\documentclass[times,twocolumn,final,authoryear]{elsarticle}

\usepackage{jasr}
\usepackage{framed,multirow}

\usepackage{amssymb}
\usepackage{latexsym}

\usepackage[bookmarks = true,colorlinks = true,linkcolor = blue,urlcolor = blue,citecolor = blue,bookmarks = true,linktocpage = true,hyperindex = true,breaklinks]{hyperref}

\usepackage{url}
\usepackage{xcolor}
\definecolor{newcolor}{rgb}{.8,.349,.1}

\usepackage{sidecap} 
\usepackage{lineno}

\newcommand{\kms}{   {km s$^{-1}$}}
\chardef\us=`\_

\journal{Advances in Space Research}

\begin{document}

\verso{Pooja Devi \textit{etal}}

\begin{frontmatter}

\title{Type II radio bursts and space weather phenomena: A statistical study} 

\author[1]{Pooja {Devi}\corref{cor1}}
\cortext[cor1]{Corresponding author:}
\ead{setiapooja.ps@gmail.com}
\author[2]{Rositsa Miteva}
\author[1]{Ramesh {Chandra}}
\author[3]{Kostadinka {Koleva}}
\author[4]{Bendict {Lawrance}}

\address[1]{Department of Physics, DSB Campus, Kumaun University, Nainital-263001, India}
\address[2]{Institute of Astronomy and National Astronomical Observatory - Bulgarian Academy of Sciences, 72 Tsarigradsko Chaussee Blvd., Sofia 1784, Bulgaria}
\address[3]{Space Research and Technology Institute, Bulgarian Academy of Sciences, Sofia 1113, Bulgaria}
\address[4]{Department of Big Data Management, Gangseo University, Seoul 07661, Republic of Korea}

\begin{abstract} 
In this paper, we present for the first time a comprehensive statistical study between type II radio bursts from the metric (m) to the dekameric$-$hectometric (DH) domain and their associated solar and space weather (SW) phenomena, namely, solar flares (SFs), sunspot (SN) configurations, filament eruptions, coronal mass ejections (CMEs), their interplanetary (IP) counterparts (ICMEs) and shocks, in situ detected particles and geomagnetic storms (GSs). The m-only and m+DH radio signatures are identified from dynamic spectra provided by the ground-based RSTN stations distributed over the globe together with Wind/WAVES satellite data. The DH-only type IIs are adopted from a ready catalog based on Wind/WAVES spacecraft data.
We perform the temporal and spatial association between the radio emission and the listed above activity events during solar cycle (SC) 24, separately for the three sub-categories, m-only, m+DH and DH-only type IIs. A quantitative assessment on the occurrence rates is presented as a function of the strength of the specific SW phenomena: highest rates are obtained with CMEs, SFs, filament eruptions, and SN configurations, whereas a much weaker relationship is found with ICMEs, IP shocks, energetic particles, and GSs. The potential of the obtained rates to be used in empirical or physics-based models for SW forecasting is discussed.
\end{abstract}

\begin{keyword}
\KWD Type II radio bursts \sep CMEs \sep Solar flares \sep Space weather
\end{keyword}
\end{frontmatter}

\section{Introduction}
\label{sec_introduction}
Radio bursts are originally classified into five different types (from I to V) based on their frequency and temporal coverage in the frequency$-$time$-$radio intensity plots \citep{1963ARA&A...1..291W}. In these, so-called dynamic radio spectra, type II radio bursts appear as one or more bands drifting from high to low frequency with slow drift rates, ranging from 100s kHz up to MHz per second, and typical duration from few to 10s of minutes \citep{1985srph.book..333N}. Although the spectra provide no information on the location of the emission over the solar disk, they can attain unprecedented temporal (ms) and frequency (kHz) resolution. The features over the dynamic spectra are then used to deduce the identity (by the burst type) and speed of the emitter (by the slope or drift), once a density model of the corona is agreed upon \citep{1947Natur.160..256P}. The emission mechanism from the metric (m) to dekametric$-$hectometric (DH) wavelengths is the plasma emission, which links the observed radio frequency to the electron density of the medium \citep{1978SoPh...57..279D,1985ARA&A..23..169D}. Together with its wide observational domain, the radio emissions in the m$-$DH wavelength range carry out a rich diagnostic potential for solar physics and space weather (SW) research \citep{Carley2020, 2021FrASS...7..105K}. The term SW refers to the influence of solar activity in the heliosphere, the planetary magnetospheres, atmospheres and surface, the effects to space-borne and ground-based technology as well as the risks to the astronaut's health and life \citep{2007LRSP....4....1P,2021LRSP...18....4T}. 

Nowadays, it is widely accepted that the type II radio signatures are due to electrons accelerated at a magnetohydrodynamic (MHD) pressure pulse or/and shock wave and finally emitting at radio frequencies \citep{Ginzburg1958,1983ApJ...267..837H,Mann2018}. Initially, the accelerated electron beams excite Langmuir waves due to beam-plasma instabilities. These waves are then converted into radio waves by nonlinear wave$-$plasma processes at the local electron plasma frequency or its harmonic(s), thus producing type II radio emissions as single or double emission bands, respectively \citep{Ginzburg1958, Thejappa2000, Frassati2019}. As far as their spectral range is concerned, the type IIs cover from the m range, e.g. about 400 MHz in the solar corona \citep{2021SoPh..296...27U}, to the DH wavelengths, e.g. down to 20 kHz in the interplanetary (IP) space, as provided by the Wind/WAVES instrument coverage \citep{1995SSRv...71..207L}. 

Historically, solar flares \citep[SFs,][]{2011SSRv..158....5H}, have been identified as the origin of type IIs \citep{1968SoPh....4...30U}. During these enormous solar explosions, however, besides the electromagnetic emission across the entire spectrum (hence the term flare), major reconstruction of magnetic field lines, mass motion and acceleration of particles takes place. With the discoveries of coronal mass ejections \citep[CMEs,][]{Tousey1973}, and the radio observations from space, type II bursts are now linked primarily with these magnetized plasma bubbles being expelled from the corona into the IP space. The CME generate the shock wave, when its speed exceeds the local Alfv\'en speed, mostly at the CME nose and flanks \citep{Cho2007, Carley2013, Zucca2014}. Although this is logical for the case of IP type IIs, the SF contribution to shocks is still a plausible scenario for the coronal type IIs.

Spectral information with high temporal resolution can be recovered even with the simplest in design radio antennas (e.g., the e-Callisto network, \url{https://www.e-callisto.org/}). However, an array of antennas is needed to provide the projection of the radio emission on the solar disk (e.g., Nan\c{c}ay Radio Heliograph \url{https://www.obs-nancay.fr/radioheliographe/}, Nobeyama Radio Heliograph \url{https://solar.nro.nao.ac.jp/norh/}, Low Frequency Array (LOFAR), \url{https://www.astron.nl/telescopes/lofar/}, and the Murchison Widefield Array (MWA), \url{https://www.mwatelescope.org/telescope/radio-astronomy/}). For this study we focus on the dynamic radio spectra only, which are plots of the radio frequency vs. time, where the radio intensity is given in a color code. 

In the MHz or also m-range, the radio emission from the Sun can be easily monitored with a variety of ground-based antenna configurations, due to the large angular size of the source and the extremely bright radio intensities emitted. There are many ground-based radio telescopes, which observe m-type II radio bursts. Data from the Radio Solar Telescope Network (RSTN) stations (25$-$180 MHz) is used for this study due to its full UT-coverage. At the lowest possible frequency limit from ground one could also use the MWA (80$-$300 MHz) \citep{Tingay2013}, NenuFAR (10$-$85 MHz) \citep{nenuFAR2012}, the Ukrainian T-shaped Radio telescope (8$-$33 MHz) 
\citep{Konovalenko2016}, and Gauribidanur Radio Interferometric Polarimeter (8$-$80 MHz) \citep{Ramesh2011}, however these radio observatories are not solar dedicated.

Radio emission at very long wavelengths (kHz-range) needs to be observed from space due to the ionospheric cutoff at the plasma frequency (of a few MHz). Traditionally, it is done by the Wind spacecraft ({\url{https://wind.nasa.gov/}}) \citep{Bougeret1995} since the 1990s, and STEREO ({\url{https://stereo.gsfc.nasa.gov/}}) \citep{Bougeret2008} since the 2010s, but recently also by the Parker Solar Probe (PSP) ({\url{http://parkersolarprobe.jhuapl.edu/}}) \citep{Fox2016} and Solar Orbiter ({\url{https://www.esa.int/Science_Exploration/Space_Science/Solar_Orbiter}}) \citep{2021A&A...646A.121G}. 
Thus, the type II counterparts observed with space-borne instruments in the IP space are known as DH type II bursts \citep[DH IIs][]{Gopalswamy2019}, or also IP IIs. There are 181 DH IIs (2009$-$2019) listed by \url{https://cdaw.gsfc.nasa.gov/CME_list/radio/waves_type2.html} whereas much more
DH IIs are listed by \url{https://spdf.gsfc.nasa.gov/pub/data/stereo/documents/websites/solar-radio/wind/data_products.html} using also data from the STEREO/WAVES mission, \citep{2008SSRv..136..529B}. 

The radio emission signatures are subject of research for about 80 years \citep{2008A&ARv..16....1P} and have been already recognized as a tool for SW forecasting \citep{2004LNP...656...49W,2018CRPhy..19...36K}. \cite{2021FrASS...7..105K} summarized the diagnostic potential of radio emission signatures, as well as their relevance for relativistic protons (observed in situ or as gamma-ray emission from the solar atmosphere) and focused on the importance of type II bursts as a proxy for shock waves and energetic protons. The latter was based on earlier findings, e.g., by \cite{Ameri2019} where it was concluded that the initial particle release is always accompanied by a m-II or DH-II burst, and the energy spectrum of the particles is harder when the initial particle release is at the time of a m-II burst or before the start of DH-II burst. 

In addition to the direct identification of plasma density, radio emissions and in particular m-to-DH type II bursts provide a magnetic field diagnostic across large distances and such radio bursts are valuable for calculating the coronal magnetic field \citep{2021FrASS...7...77A}. This is a valuable estimation for the magnetic fields (strength and direction) which can be used to constrain models of magnetic field extrapolation.

A critical overview on the applicability of type II signatures to CME observations (i.e., shock wave tracking, SEP generation, magnetic field orientation in the ejecta) was provided by \cite{2020FrASS...7...43V} stressing on the importance of spectropolarimetric radio imaging data. Since type II radio bursts are the radio signatures of shock waves (and thus relevant to CMEs and/or SFs), it is important to relate them to other shock signatures (e.g., observed in different parts of the electromagnetic domain, if possible) or/and to other agents of shock-associated solar activity, e.g., filament eruptions, ICMEs, IP shocks, energetic particles and magnetospheric disturbances. In addition, we explore the predominant type of magnetic configuration of the underlying sunspots and the strength and location of the related SFs. 

Below we describe in short the main features of the solar and SW phenomena that will be in the focus of this study. The correlations between the SW phenomena to the type IIs are provided when found in the literature. The opposite direction for the correlation, namely starting with a list of type IIs and calculating the occurrence rates between type IIs and a given SW event, is the main goal of our work and is presented in Section~\ref{res}. 

\begin{itemize}
\item {\bf SFs}

SFs have been observed with several generations of GOES spacecraft since the late 1960s (\url{ftp://ftp.swpc.noaa.gov/pub/warehouse/}). A well-known classifier for SFs is according to the peak in soft X-rays (SXRs), termed A (the weakest), B, C, M and X-class (the extreme ones). Recent studies on X ($\geq$10$^{-4}$ W m$^{-2}$) and M-class (10 times less intense) SFs can be used to calculate the overall association rates with type IIs. 
Based on the list of 49 X-class SFs in solar cycle (SC) 24, created by \cite{2021BlgAJ..35...87M}, we performed the respective associations and calculated the occurrence rates in SC24: an X-class SF has 67\% chance to be related to m IIs and in 45\% it is associated with DH II 
(\url{https://cdaw.gsfc.nasa.gov/CME_list/radio/waves_type2.html}). For M-class SFs \cite{Miteva2022a} reports  about 7\% association with DH IIs in SC24 (when the largest M-class SFs are considered, the fraction increases).

\item {\bf SN configuration}

The sunspot (SN) number is the oldest database available in the solar physics domain. For this work, we adopt the monthly numbers reported by \url{https://www.sidc.be/SILSO/datafiles}. Furthermore, the Mount Wilson Sunspot Magnetic Classification, \url{https://www.spaceweather.com/glossary/magneticclasses.html}, is used throughout this study. We are not aware of any statistical associations between the given types of sunspots and radio emission signatures.

\item {\bf CMEs} 

The information about the CMEs, e.g., time of first appearance, projected speed, angular width (AW), measurement position angle (MPA), is often taken from the CDAW SOHO LASCO CME catalog, \url{https://cdaw.gsfc.nasa.gov/CME_list/}, which is the source used by us as well. Alternative CME catalogs can be found at \url{https://cdaw.gsfc.nasa.gov/CME_list/links.html}. Despite the noted limitations of CME identifications by the manual compared to the automatic catalogs \citep{Yashiro2008}, the use of the manual one \citep{GopCDAW2009} by the research community is prevalent. In SCs 23 and 24, there are about 30 000 CMEs being reported by CDAW, about half of them in SC24. The same team reports the IP IIs with CME signatures, which are about 520 for both SCs, \url{https://cdaw.gsfc.nasa.gov/CME_list/radio/waves_type2.html}.
A recent study by \citet{Kumari2023} reports 3\% for the association between CMEs and m IIs in SCs 23 and 24. There are several studies in literature where the association between CMEs and type II radio bursts (both m and DH) is investigated in detail \citep[][and references cited therein]{Kahler1986, Gopalswamy2012, Gopalswamy2015, Kharayat2021}.

\item {\bf Filaments}

Filaments are elongated arcade-like structures of denser and colder plasma compared to the surrounding solar corona when viewed over the solar disk \citep{2014LRSP...11....1P}. They are located above the polarity inversion line in the photosphere and thus related to the local magnetic field. The term `prominences' is used when the same phenomena is visible above the solar limb. Erupting filaments have been associated with CMEs, however, the rate varies from just a few, to 50$-$60\%, $\sim$70\% \citep{2015SoPh..290.1703M} or up to 90\%, see also discussion in \cite{2014LRSP...11....1P} for earlier works. Despite the large disagreement, at least some of them can be regarded as a driver of shock waves. Thus, the causal relationship between filaments and type II radio bursts can be explored via statistical means. Several online lists of erupting filaments with partial time coverage are know to us: 2010$-$2014 \url{https://aia.cfa.harvard.edu/filament/}, 2016$-$2019 \url{https://www.kwasan.kyoto-u.ac.jp/observation/event/sddi-catalogue/}, 2010$-$2020 \url{https://cdaw.gsfc.nasa.gov/CME_list/autope/}. Based on the latter (automatic) catalog of 1047 filaments in SC24, the ratio with the m-type II radio bursts amounts to 44\% which should be regarded as an upper limit only. Similarly, the ratio with DH-only IIs amounts to 8\%.

\item {\bf ICMEs}

The IP manifestation of CMEs, termed ICMEs, are usually studied by in situ measurements of plasma parameters and magnetic field components at 1 AU \citep{2006SoPh..239..393J}. A different set of parameters and criteria are used in the literature to define the boundaries of these structures and several sub-types exist, which are not relevant for our kind of study. Online catalogs of ICMEs, with their timing, parameters and solar origin are available, e.g., \url{https://izw1.caltech.edu/ACE/ASC/DATA/level3/icmetable2.htm} and \url{https://wind.nasa.gov/ICME_catalog/ICME_catalog_viewer.php}, depending on the spacecraft of observation. A number of parameters have been identified and provided. Due to their IP essence, ICMEs are usually associated with DH IIs. A recent study on the relationship between ICMEs and DH IIs was completed by \cite{Patel2022} and reported $\sim$47\% association rate in SC24.

\item {\bf IP shocks}

Abrupt jump in the solar wind speed and plasma parameters are used to define the arrival of a shock wave at the spacecraft (usually at 1 AU). Based on the physical nature of type IIs and IP shocks, we would expect a high association between the two phenomena. A simple ratio between the IP shocks and DH IIs gives an upper limit for the occurrence rate of 95\%. A reliable online source for IP shocks is provided by \url{https://lweb.cfa.harvard.edu/shocks/} that reports 191 IP shocks by the Wind spacecraft (2009$-$2019). 

\item {\bf In situ particles}

Since one of the possible acceleration mechanisms of solar particles is at the shock wave fronts, one would expect a positive correlation between the particles and the shock signatures. The most common species observed in situ are 10s MeV protons and 100s keV electrons. \cite{Miteva2017} presented the association between solar energetic particles (\url{http://www.stil.bas.bg/SEPcatalog/}) and type II bursts in SC24: 59\% for m, and 73\% for DH IIs. A detailed review on previous works can be found in \cite{2021FrASS...7..105K,Klein2021}. \citet{Ameri2019} analysed the 58 proton events in energy range 55$-$80 MeV from 1997$-$2015 and look into their relation with m and DH-type II radio bursts. They found that about 19\% of proton events were associated with m-type II radio bursts, which is little lower value probably due to different selection criteria and time period, as well as the smaller event sample under the study.
More accurately, \cite{Miteva2022} reported for the first time on the association between SEEs, \url{https://www.nriag.sci.eg/ace_electron_catalog/} \citep{Samwel2021} and the (electron-generated) radio emissions. When starting with the list of in situ electrons, the association rate with m IIs in SC24 is about 25\%, whereas with DH IIs reaches up to 29\%.

\item {\bf GSs} 

It is well-established that the primary driver of major geomagnetic storms (GSs), \cite{1961PhRvL...6...47D}, are the ICMEs with southern component on their magnetic field, so-called $B_z$ \citep{2012JSWSC...2A..01R,2015RAA....15...85R}. Weaker GSs could be also driven by fast streams of solar wind and/or co-rotating or streaming interaction regions (CIRs or SIRs). Again, due to the IP source of both phenomena, a physical relationship could be sought between GSs and DH IIs (regarded as the radio signatures of ICMEs) by means of statistical association. A list of strong GSs (defined as disturbances with the Dst index $\leq -100$~nT) can be found in \cite{2023AdSpR..72.3440S} and its extension for Dst index $\leq -50$~nT in \cite{2023Atmos..14.1744M} (see also \url{https://catalogs.astro.bas.bg/}). Based on the latter list, we calculated the association rate in SC24 for the 29 strong ($-100$ nT) GSs to be about 38\% with DH IIs, whereas for the extended list of 185 GSs ($-50$ nT) the results are 19\% with DH IIs.

\end{itemize}

There is a rich volume of literature on the single case and statistical studies of type II radio bursts, however a historical overview goes beyond the scope of our work. Below, we aim to highlight some statistical results.

Comprehensive statistics over long periods depend on the instrument data coverage. With the launch of Wind satellite, dynamic radio spectra in the DH (up to km) range were provided on routine basis leading to the compilation of catalogs and correlation studies with different phenomena \citep{2000GeoRL..27.1427G,2019SunGe..14..111G}. Another study on DH type IIs over SCs 23 and 24 \citep{Patel2021} reported that the CMEs with higher speeds are associated with type II bursts displaying extended frequency emissions. In addition the speed and width of the CMEs increases from m to m--km type II bursts implying a progressive increase in kinetic energy, consistent with earlier reports \citep{Gopalswamy2010}. Geoeffective CMEs were found to be with or without DH type II emission \citep{Patel2022} and the type II associated CMEs are faster in comparison to non-type II associated CMEs.

Another aspect of the reported statistical studies in the past is the direction of association. Namely, one could start from a list of solar energetic particle events \citep{Miteva2017,Miteva2022} or SFs \citep{Miteva2022a} and then search for radio emission signatures. In the former works, the electron-association rates with type IIs are used to differentiate between flare (up to 29\%), CME (18\%) or mixed (17\%) origin, whereas the latter study show the weak association rate (7\%) between M-class flares in SCs 23 and 24 and type IIs. In contrast, the analysis in this manuscript provides the other direction of association, namely we started from a catalog of type II radio bursts \citep{Lawrance2024} and calculated the association rates with a wide range of phenomena.

Our study starts similarly to the analyses completed by \citet{Kumari2023}, using m-type IIs. Also, \citet{Kumari2023} found 95\% of m IIs to be associated with CMEs, and only 3\% of CMEs to be associated with m II bursts. However, these authors focused exclusively on the relationship between the already provided observatory reports of m-IIs from \url{https://www.swpc.noaa.gov/products/solar-and-geophysical-event-reports} and CMEs from various databases for the last two SCs. In contrast, we consider the associations with different solar and SW phenomena as detected during SC24 (2009$-$2019). For the same period of our analysis, they found very similar number of m IIs in the observatory reports (435), compared to the initial catalog of m IIs (429). Then we made a strict distinction into m-only (342) and m+DH IIs (87), both based on visual identifications. Furthermore, we complemented the original list \citep{Lawrance2024} by (89) DH-only II signatures.

In this paper, we aim to investigate the associations between, the three sub-types of type II bursts: m-only, m+DH, and DH-only IIs on one side, and a variety of solar and SW phenomena on another, during SC24. A quantitative assessment on these associations is performed by means of calculating the occurrence rates (e.g., the fraction of the type IIs associated with a particular SW event reported in percentage), trends (dependencies with respect to a given physical parameter) and overall properties (in terms of mean, median or peak values). 
In contrast to previous works, the current study identifies, on one side, the occurrence rates of type II bursts - in its entirety and in several wavelength sub-divisions: m-only/m+DH/DH-only IIs - with a wide variety of phenomena, e.g., SN configuration, SFs, CMEs, filaments, ICMEs, IP shocks, in situ particles and GSs and, on another, provides a systematic description on the trends that are representative over the entire SC24.

The paper is organized as follows: Section \ref{data} provides information about the instruments and the data they provide, the association methods applied, and other adopted procedures. Section \ref{res} presents the results of the m-only/m+DH/DH-only IIs and the related SW phenomena, as described above. A discussions on the results is given in Section \ref{discussion}. Finally, Section \ref{summary} summarizes the main outcomes of the study.

\section{Data sources and association procedures} 
\label{data}

We base our analysis on the catalog of m-type II radio bursts in SC24 (2009$-$2019) compiled from the dynamic spectral data provided by the RSTN \citep{Lawrance2024}. The RSTN is a network of four identical solar observatories over the globe allowing for (nearly) continuous observation in time. Namely, quick-look dynamic spectrum data from the RSTN with frequency coverage from 25 to 180 MHz have been used: \url{https://www.ngdc.noaa.gov/stp/space-weather/solar-data/solar-features/solar-radio/rstn-spectral/}. The quick-look plots were visually inspected for type II signatures. A freely-accessible version of the m II catalog will be supported online at \url{https://catalogs.astro.bas.bg/}. The initial list of m-type II bursts was further complemented here by adding their DH counterparts by the Wind/WAVES. Then the list is split into m-only II category (342 cases), in case no DH type II can be identified, and m+DH IIs (87 cases) in case of a occurrence of type IIs in the m and DH-ranges. Finally, the DH-only II category consists of events when no m IIs is identified at the same time (89 cases). Thus, the final type II event list used in our study amounts to 518. The times of occurrence and associated SFs/CMEs of the latter category are adopted from the event list provided by \url{https://cdaw.gsfc.nasa.gov/CME_list/radio/waves_type2.html}.

For the purpose of this study, the onset times of the m IIs, as well as their associated solar origin, in terms of SFs and CMEs are used. The association is done based on timing (up to 1-hr difference between m II with SF/CME occurrences) and location criteria (the SF longitude/latitude and the MPA of the CME to be in same solar quadrant for the given event). The SF and CME timings and physical parameters are adopted from the readily provided catalogs: \url{ftp://ftp.swpc.noaa.gov/pub/warehouse/} and \url{https://www.solarmonitor.org/} for the SFs and \url{https://cdaw.gsfc.nasa.gov/CME_list/} for the CMEs. Additional information about the active region (AR) or SN classification is collected from the daily reports prepared by the NOAA/Space Weather Prediction Center: \url{https://ngdc.noaa.gov/stp/spaceweather.html}.

The DH-counterparts of the m IIs (for the m+DH category) are  identified using the dynamic radio spectra as provided by the following data sources:
\begin{itemize}
\item \url{https://cdaw.gsfc.nasa.gov/CME_list/radio/waves_type2.html}
\item \url{https://solar-radio.gsfc.nasa.gov/wind/data_products.html}
\item \url{https://secchirh.obspm.fr/}
\end{itemize}
The association between the m and DH IIs is done based on their mutual timing. For the purpose of this work we do not discriminate between SF vs. CME-driven m-type II bursts. 

Published lists of in situ detected solar energetic particles are used for both protons (\cite{Miteva2018} from the Wind/EPACT instrument) and electrons (\cite{Samwel2021}, from ACE/ EPAM instrument, respectively). The association between m-only/m+DH/DH-only and the SEPs is done based on the requirement both phenomena to share the same solar origin, e.g. SF and/or CME.

Several data sources are used for the identification of filaments, namely:
\begin{itemize}
\item \url{http://solar.nro.nao.ac.jp/norh/html/prominence/}
\item \url{https://cdaw.gsfc.nasa.gov/CME_list/autope/}
\item \url{https://aia.cfa.harvard.edu/filament/}
\item \url{https://www.lmsal.com/isolsearch}
\end{itemize}
and the association between the m-only/m+DH/DH-only IIs and filaments is done via their accompanied CME. 

For the identification of potential IP and magnetospheric signatures of m II we used the requirements for the associated CME in either case to be the same ejecta and the transport time from the Sun to Earth to correspond to the time of occurrence of ICMEs, IP shocks, and GSs. In the former two cases, we have information from both Wind and ACE spacecrafts and thus for the list of ICME events we used:
\begin{itemize}
\item \url{https://wind.nasa.gov/ICME_catalog/ICME_catalog_viewer.php}
\item \url{https://izw1.caltech.edu/ACE/ASC/DATA/level3/icmetable2.htm},
\end{itemize}
respectively, whereas, for IP shocks we used the following databases:
\begin{itemize}
\item \url{http://www.ipshocks.fi/database}; 
\item \url{https://lweb.cfa.harvard.edu/shocks/wi_data/}.
\end{itemize}

Again, the CME-origin of m-only/m+DH/DH-only IIs is used to perform their association to GSs. As a representative value for the GS strength, we use the disturbance storm index (Dst), measured in nT, as negative values. Hourly reports of the Dst index are available by the Kyoto database, in their final (for the period 2009$-$2016) or provisional (2017$-$2019) form, whereas the solar origin of the GSs is adopted from \cite{2023Atmos..14.1744M}. The respective links to these databases are:
\begin{itemize}
\item \url{https://wdc.kugi.kyoto-u.ac.jp/dstdir/index.html}; \item \url{https://wdc.kugi.kyoto-u.ac.jp/dst_provisional/index.html};
\item \url{https://catalogs.astro.bas.bg/}.
\end{itemize}

As we used ready catalogs of solar and SW phenomena, we rely on their intrinsic accuracy and completeness. Manual or automatic, a catalog follows a prescribed set of criteria for its compilation, which is a time consuming effort. Based also on our experience we accept that the inherent uncertainty of a given event list depends on data availability (e.g., coverage, gaps, instrument sensitivity) and the selection criteria employed. The subjectivity plays a larger role for the manual catalogs, whereas erroneous identifications are often the case for automatic ones. Thus, the reported below rates and trends are based, on one hand, on the adopted by us association techniques and on the reliability of the used by us catalogs, on the other. Erroneous cases could well be present, however we do not expect their number to significantly influence the reported results.

Since in this study we focus on occurrence (association) rates, i.e. the number of observed events normalized to a given event sample, we use the formula of the propagation error of a ratio (e.g., \url{https://www.geol.lsu.edu/jlorenzo/geophysics/uncertainties/Uncertaintiespart2.html}). There, the absolute errors of the two samples are calculated as the square root of each sample size.

In brief, the methodology is summarised as follows: Step 1: We start with the catalog of m-type II radio bursts from RSTN instruments provided by \cite{Lawrance2024}. Step 2: This m-type II list is further complemented by DH counterparts from the data of Wind/WAVES and, m-only, m+DH, and DH-only type II categories have been formed. Step 3: The association of these type II radio bursts with different solar (SFs, filament eruptions, and CMEs) and space weather (ICMEs, IP shock, GSs, and in-situ particles) events is based on the information provided in available catalogs and applying a set of association procedures, as described above.

\section{Results}
\label{res}

The following color-notation is used in the histograms henceforth. The type IIs with a coronal origin are plotted together in a stacked histogram. Namely, the grey-colored portion of the bars represent the m-only IIs, whereas the black parts highlight the m+DH type II bursts. Next to them, as separate bars in red color, are plotted the DH-only type IIs. The numbers of all three types of type IIs are given as \%, calculated as a ratio from the total number of type II events (518). The trends of the type II radio bursts and their relation with the other solar and SW phenomena are organized in the following subsections. 

\subsection{Type II bursts in the m/DH range}

The occurrence rate of m-only IIs is estimated to be 66\% (342/518), 17\% (87/518) are m+DH IIs, and the remaining 17\% (89/518) are DH-only IIs. Note that the physical relationship for the m+DH type II category is only implied here, based on the co-occurrence of type IIs in the m and DH range within the time window of interest. By comparing m-only and m+DH IIs, it is evident that there are fewer number of m+DH IIs, which suggests that most of the type IIs in the m-range are not necessarily accompanied by DH II signatures.

\begin{figure}[t]
\includegraphics[width=\columnwidth]{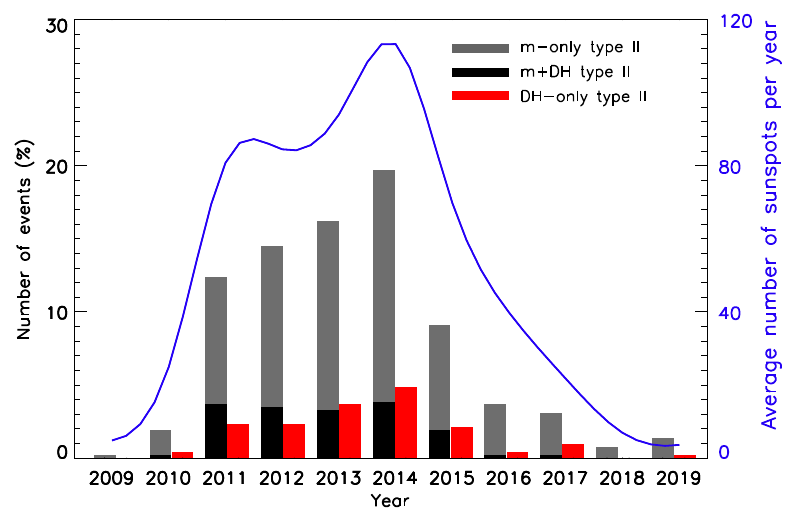}
    \caption{Annual distribution of m-only/m+DH/DH-only type II radio bursts for SC24, denoted with grey, black and red bars, respectively. The blue curve tracks the averaged number of sunspots per year.}
    \label{fig:year}
\end{figure}

The annual distributions of type IIs and sunspots are shown in Figure~\ref{fig:year}. The number of type II radio bursts in the m-range only (the grey-colored part of each bar) follows the overall trend of the number of SNs in SC24 (shown with the blue curve). The m+DH IIs however show a flat trend in the rising phase of SC24 and almost no occurrences in the declining phase. Their trend is similar to that of the DH-only IIs (red color bars) which is slightly peaked at the SN maximum in 2014.

\subsection{Type II bursts and solar phenomena}

In this Section we explore the association rates between solar phenomena and the sub-categories of type II events. A summary of the results is shown in Table~\ref{tab-solar}. The distributions of the association rates are given in detail in the respective subsections below.

\begin{table}[t]
\caption{Table for the percentage of the solar phenomena associated with m-only/m+DH/DH-only II radio bursts given in \%, calculated as the ratio to entire list of 518 type IIs. In parenthesis are given the exact number of events in each category.}
\begin{tabular}{cccc}
\hline
\setlength{\tabcolsep}{0cm}
Solar &  \multicolumn{3}{c}{Association with type II} \\
phenomena & m-only & m+DH & DH-only\\
\hline
SF class &  \\
C-class   &  24\% (123)  &  2\% (11)  & 4\% (23) \\ 
M-class   &  16\% (84)  &  9\% (48)  & 4\% (22) \\
X-class   &  2\% (12)  &  4\% (21)  & 0.8\% (4) \\
\hline
SF impulsiveness & \\ 
1--15 min &  33\% (175)  &  5\% (28)  & 2\% (13) \\
$>15$ min &  12\% (63)  &  10\% (52)  & 7\% (36) \\
\hline
CMEs  &   \\
$\le500$ \kms    &  28\% (143)  &  2\% (11)  & 2\% (9) \\ 
501--1000 \kms &  14\% (71)  &  5\% (26)  & 7\% (39) \\ 
$>$1000 \kms   &  3\% (17)  &  9\% (48)  & 7\% (35) \\ 
non-halo       &  38\% (199)  &  4\% (22)  & 7\% (38) \\
halo           &  6\% (32)  & 12\% (63)  & 9\% (45)  \\ 
\hline
Filaments &  25\% (132)  &  11\% (57)  & 8\% (41) \\
\hline
\end{tabular}
\label{tab-solar}       
\end{table}

We explicitly calculate the propagation of the error (or uncertainty) only for the largest and smallest value in the table: the occurrence rate of m-only IIs with non-halo CMEs is obtained to be 38$\pm$3\%, whereas the DH-only II association with X-class flares is 0.8$\pm$0.4\%. Thus, as the number of type IIs in the given sub-category drops, the uncertainty increases. Although the statistical significance of the smallest event samples needs to be considered with caution, the m-only II group tends to be statistically different from the remaining two sub-samples.

\subsubsection{Type II bursts and SFs}

In this study, we identified around 71\% (367/518) SFs which are regarded as associated to the type IIs, according to the criteria adopted by us. For the remaining 29\%, we either have no SF candidate (or alternatively the SF could be behind the soalr limb) or we could not identify a SF with certainty. The numbers of SFs associated with m-only/m+DH/DH-only IIs are 46\% (238/518)/15\% (80/518)/9\% (49/518), respectively.  

\begin{figure}[t]
\includegraphics[width=0.9\columnwidth]{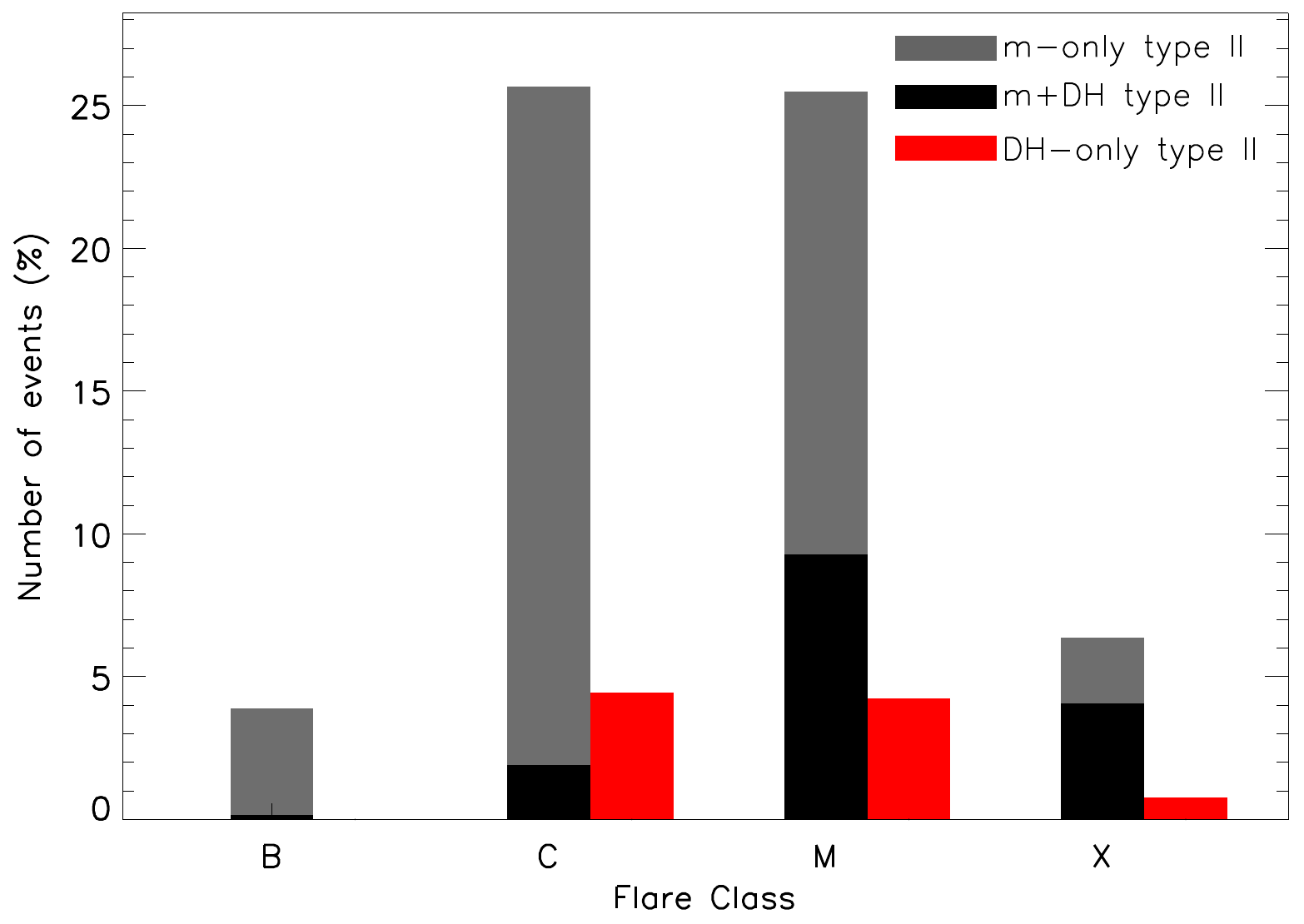}
    \caption{GOES SXR class distribution of the SFs associated with m-only/m+DH/DH-only type II radio bursts. Color-code as in Figure~\ref{fig:year}.}
    \label{fig:SF}
\end{figure}

\begin{figure}[t]
	\includegraphics[width=0.9\columnwidth]{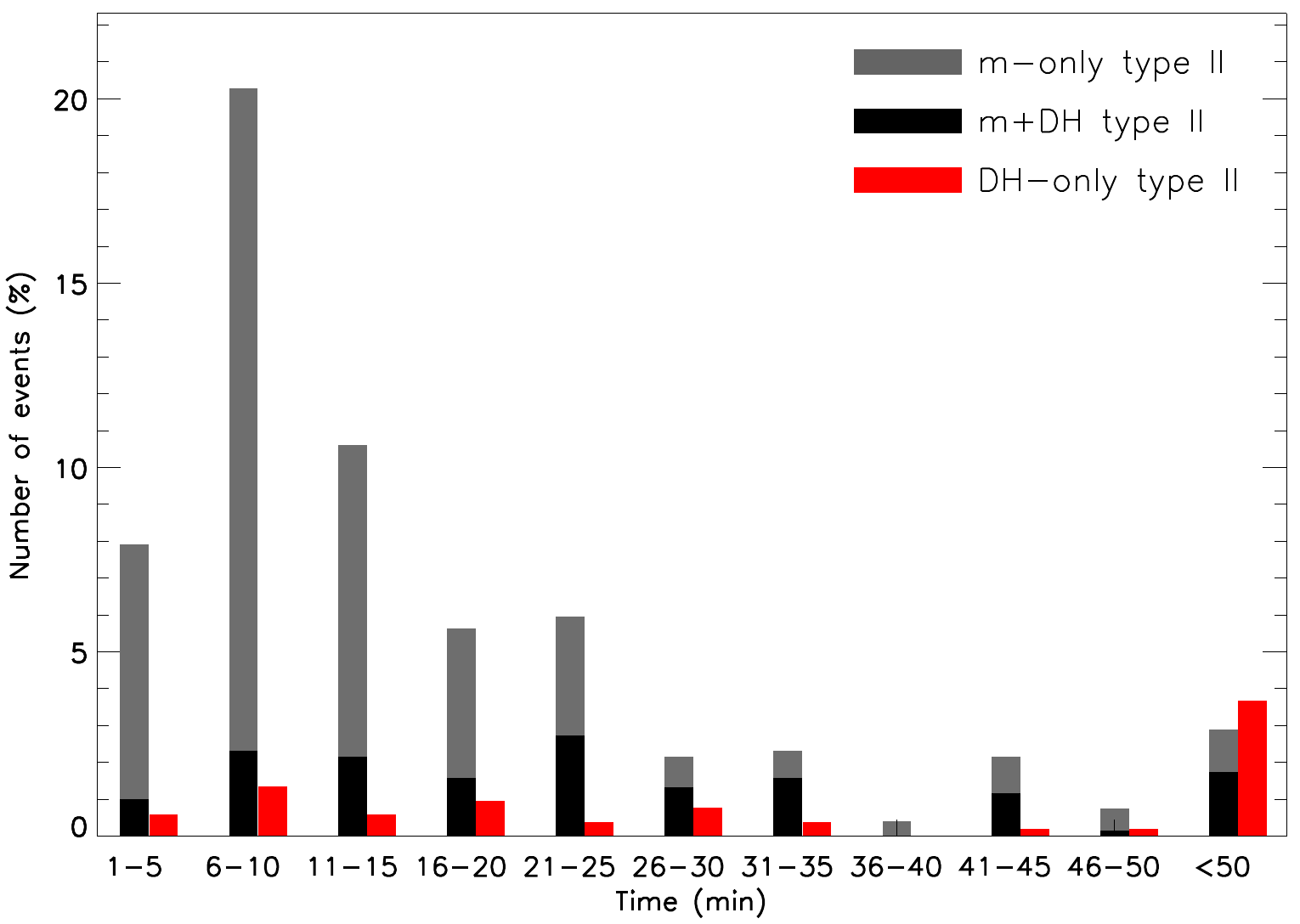}
    \caption{Distribution of the impulsiveness of SFs associated with m-only/m+DH/DH-only type II radio bursts. Color-code as in Figure~\ref{fig:year}.}
    \label{fig:SFim}
\end{figure}

\begin{figure}[t]
	\includegraphics[width=0.9\columnwidth]{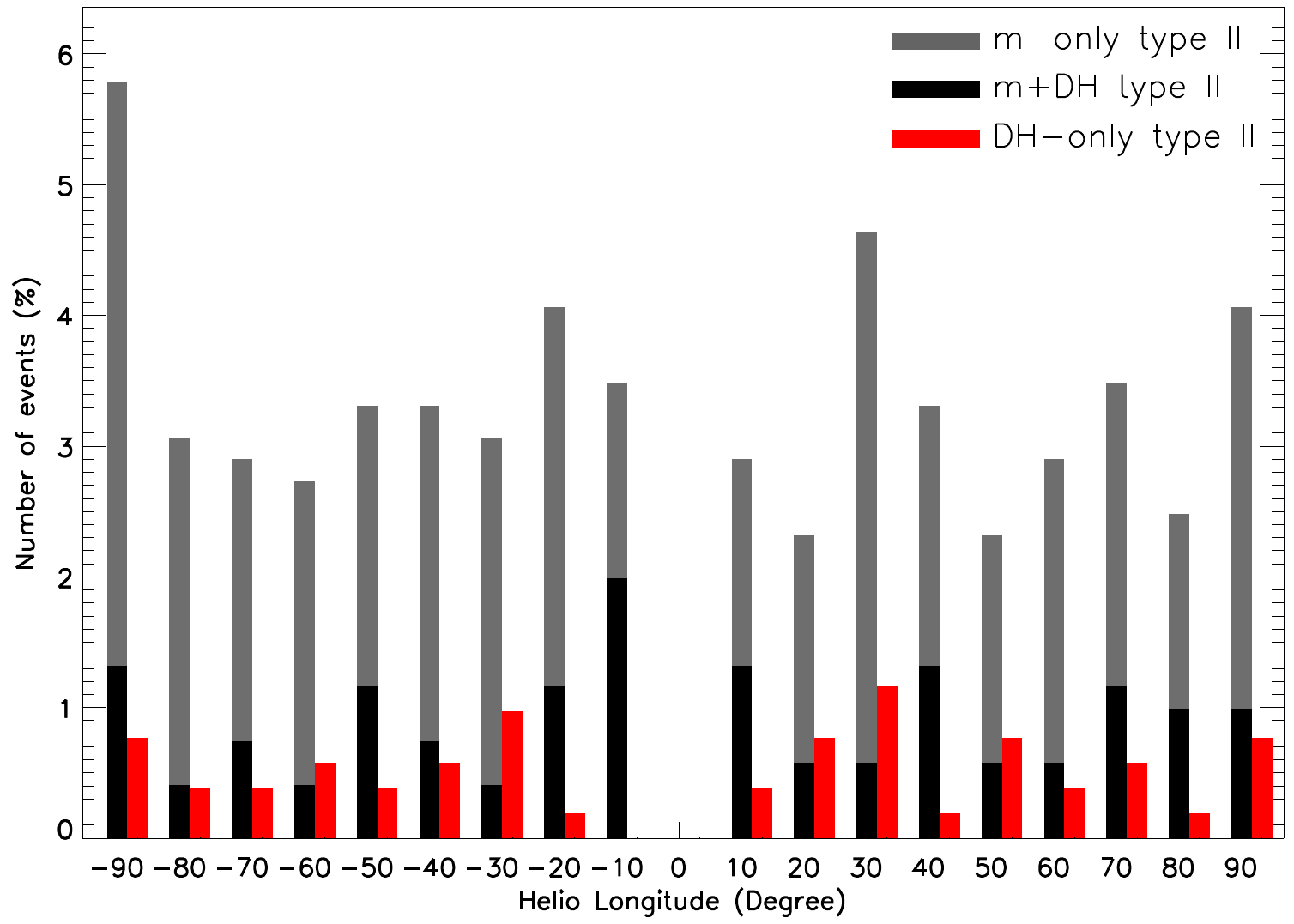}
    \caption{Distribution of the helio-longitudes of SFs associated with m-only/m+DH/DH-only type II radio bursts. Here, $-10$ corresponds to longitudes from $-10$ to 0, $-20$ for $-20$ to $-10$, etc. (East), and 10 for 0 to 10, 20 for 10 to 20, and so on (West). Color-code as in Figure~\ref{fig:year}.}
    \label{fig:SFlon}
\end{figure}

\begin{figure}[h!]
	\includegraphics[width=0.9\columnwidth]{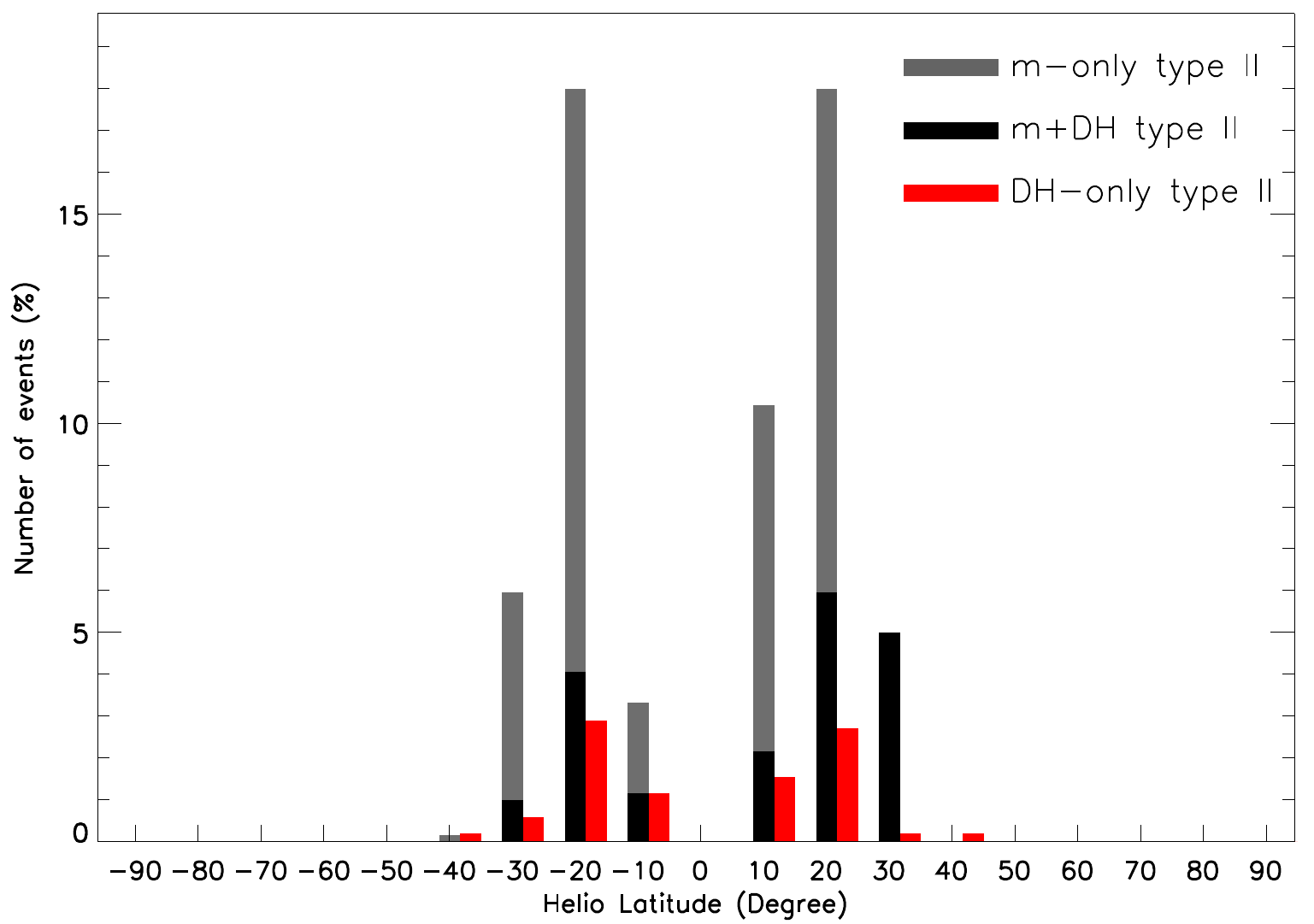}
    \caption{Distribution of the helio-latitudes of SFs associated with m-only/m+DH/DH-only type II radio bursts. The bin size for the bars are same as in Figure~\ref{fig:SFlon}. Positive values are for North, negative for South. Color-code as in Figure~\ref{fig:year}.}
    \label{fig:SFlat}
\end{figure}

Now, we explore in more details the behaviour of SFs associated type II bursts, using the same three categories as described in the above subsection. Namely, we plotted the number of m-only/m+DH/DH-only IIs associated SFs as a function of their soft X-ray (SXR) class, impulsiveness, longitude, and latitude, see Figures~\ref{fig:SF}$-$\ref{fig:SFlat}, respectively.

Figure~\ref{fig:SF} shows the distribution of the SF-associated type IIs as a function of the SXR intensity or flare class. The m-only II events exhibit the highest occurrence rate with C class SFs (up to 25\%), less with M-class SFs (up to 15\%) whereas the association with B and X class SFs is much less pronounced. On the other hand, m+DH II events show the highest association with M class SFs (about 10\%) and much lower with X and C-class SFs. The distribution of SFs associated with DH-only IIs is flat for C and M-class SFs with insignificant contribution of X-class SFs. The distributions of the m+DH and DH-only IIs are shifted towards larger SFs compared to the m-only IIs predominantly associated with C (and M)-class SFs.

Only a minority, about 2\% (12/518), of the X-class SFs are m-only IIs, which constitutes an example of strong solar eruption that produced coronal shock waves only. Among them only 2/12 can be regarded as confined, as no clear CME signatures are observed. The X-class SFs related to m+DH IIs are 4\% (21/518), whereas those associated with DH-only IIs are only 4 events. Despite the low number of X-class SFs driving type II events, the strength of the solar eruption facilitates the IP transition of coronal phenomena, as in 25 out of 37 cases, the type II bursts driven by X-class SFs reaches the DH range.

Figure~\ref{fig:SFim} illustrates the distribution of impulsiveness among SFs associated type IIs. The impulsiveness is defined as the time difference between the flare peak and its onset. The analysis reveals that the majority of SF-associated m-only IIs, i.e., 18\% (93/518), are impulsive, with rise times of the SXR emission within the range of 6 to 10 minutes, followed by 9\% (45/518) of 11$-$15 minutes and 7\% (36/518) covering the shortest periods of 1$-$5 minutes. On the other hand, SFs linked to m+DH IIs do not demonstrate a preferred duration, covering a broad range in time, from 5 to over 50 minutes, whereas the DH-only IIs are preferentially related to long-duration SFs (up to 50 min).

The longitudinal and latitudinal distribution of all three categories of type IIs are depicted in Figures ~\ref{fig:SFlon} and \ref{fig:SFlat}, respectively. Based on the histograms, it becomes evident that there is no discernible trend in terms of east and west longitudes for m-only IIs. A slightly pronounced bin at the 10 degree in the Northern hemisphere is noticed for the m-only IIs, whereas DH-only IIs tend to peak at 20 degrees in both hemispheres. The helio-latitudes for all three groups of IIs are clustered around the AR belts with a peak at $\pm20$ degrees. The latter results are in agreement with \citet{Gopalswamy2019} who found the same pattern for DH type II radio bursts during SCs 23 and 24.

\subsubsection{Type II bursts and CMEs}

Similarly, we investigate the parameters of the associated CMEs, focusing on the CME linear speed and AW within SOHO/LASCO field of view (FOV). Note that, the provided CME speeds are not the true ones as they are projected speeds. \citet{Balmaceda2018} showed that in general there is an underestimation of 20\% of the projected speeds. However, it has been recently shown \citep{Verbeke2023,2023Univ....9..179M} that the process of CME de-projection is still a very subjective procedure, depending on multiple parameters within the selected fitting model (e.g., spherical, elliptical, graduated cylindrical shell). Since there are no catalogs of 3D CME speeds in SC24 with uncertainties depending on the methodology used, we used the projected CME speed for our sample. Moreover, projected speeds are still widely used in solar physics research.

The number of all type II events associated with CMEs is 77\% (399/518), which is consistent with \citet{Kumari2023} for SC23 where they found 79\% association, however, they report this association to be 95\% in SC24. In our list, among the 399 CME-associated type IIs, m-only are 45\% (231/518) as a fraction of the entire type II list or 58\% (231/399) as a fraction of the CME-associated sample. The m+DH II events are only 16\% (85/518), or 21\% (85/399), similar to the DH-only II fraction of 16\% (83/518). The results are similar to those with the SFs for the m-only and m+DH, wheres the fraction of CME-associated DH-only IIs is nearly twice as large compared to the 9\% of SF-associated DH-only IIs. 

\begin{figure}[t]
	\includegraphics[width=0.9\columnwidth]{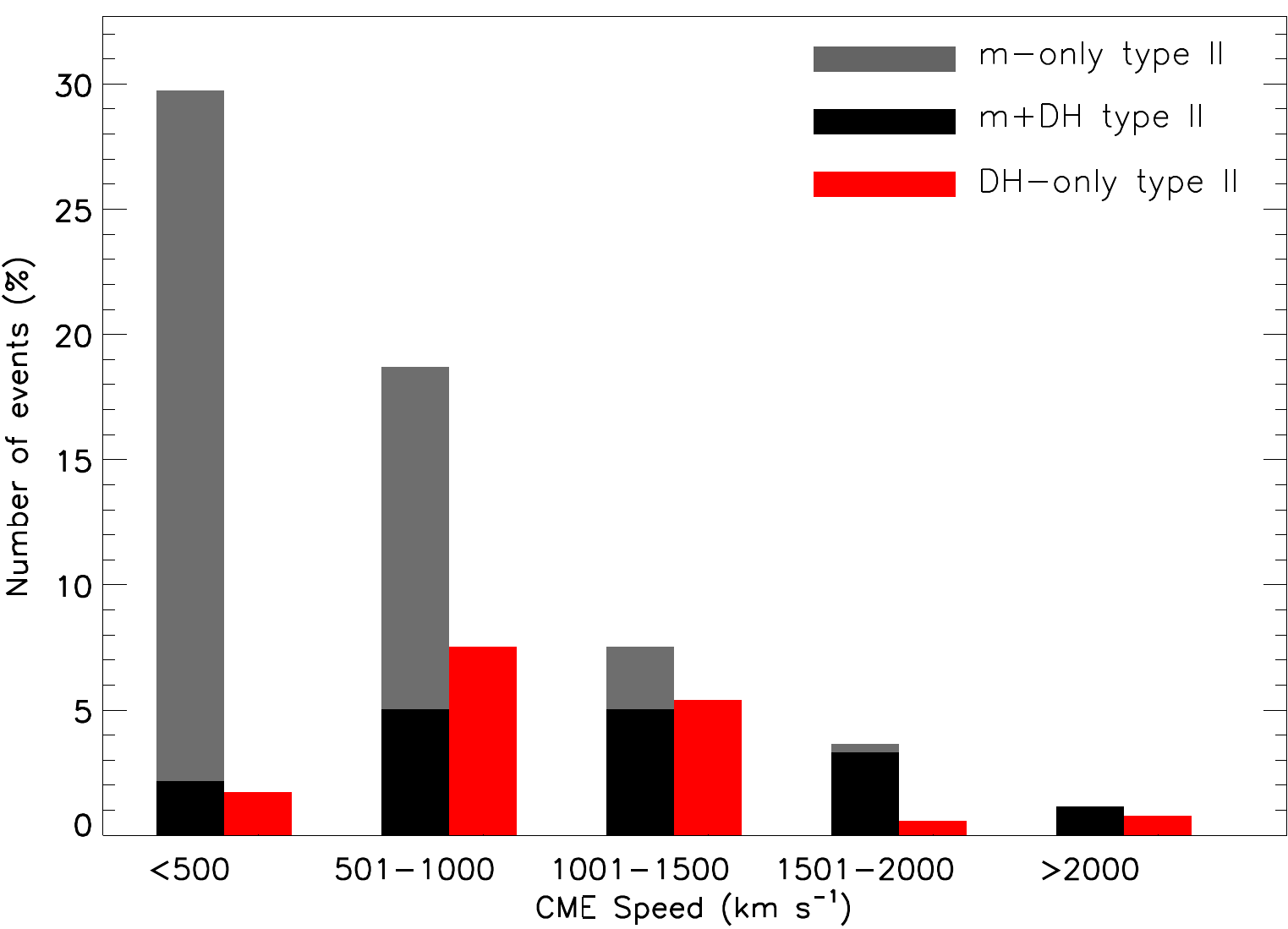}
    \caption{Histogram of the average speed of the CMEs associated with m-only/m+DH/DH-only type II radio bursts. Color-code as in Figure~\ref{fig:year}.}
    \label{fig:CME-sp}
\end{figure}

When the above associations are inspected as function of the CME speed, we note that the majority of the m-only IIs are associated with CMEs $\le$ 500 \kms~whereas the m+DH and DH-only II events are associated mostly with the CMEs having speed in the range of 501$-$1500 \kms~(see Figure \ref{fig:CME-sp}). This is consistent with earlier reports by \cite{Gopalswamy2015CME} who found that the CMEs with DH IIs have speed greater than 947 and 528 \kms~ during the two peaks of SC24. Moreover, the CMEs with speed greater than 1500 \kms, are associated exclusively with m+DH and/or DH-only IIs. This confirms the earlier results that m-only IIs associated with faster CMEs extend to DH-wavelengths \citep{Gopal2002,Prakash2014,Gopalswamy2015CME}.

\begin{figure}[t]
	\includegraphics[width=0.9\columnwidth]{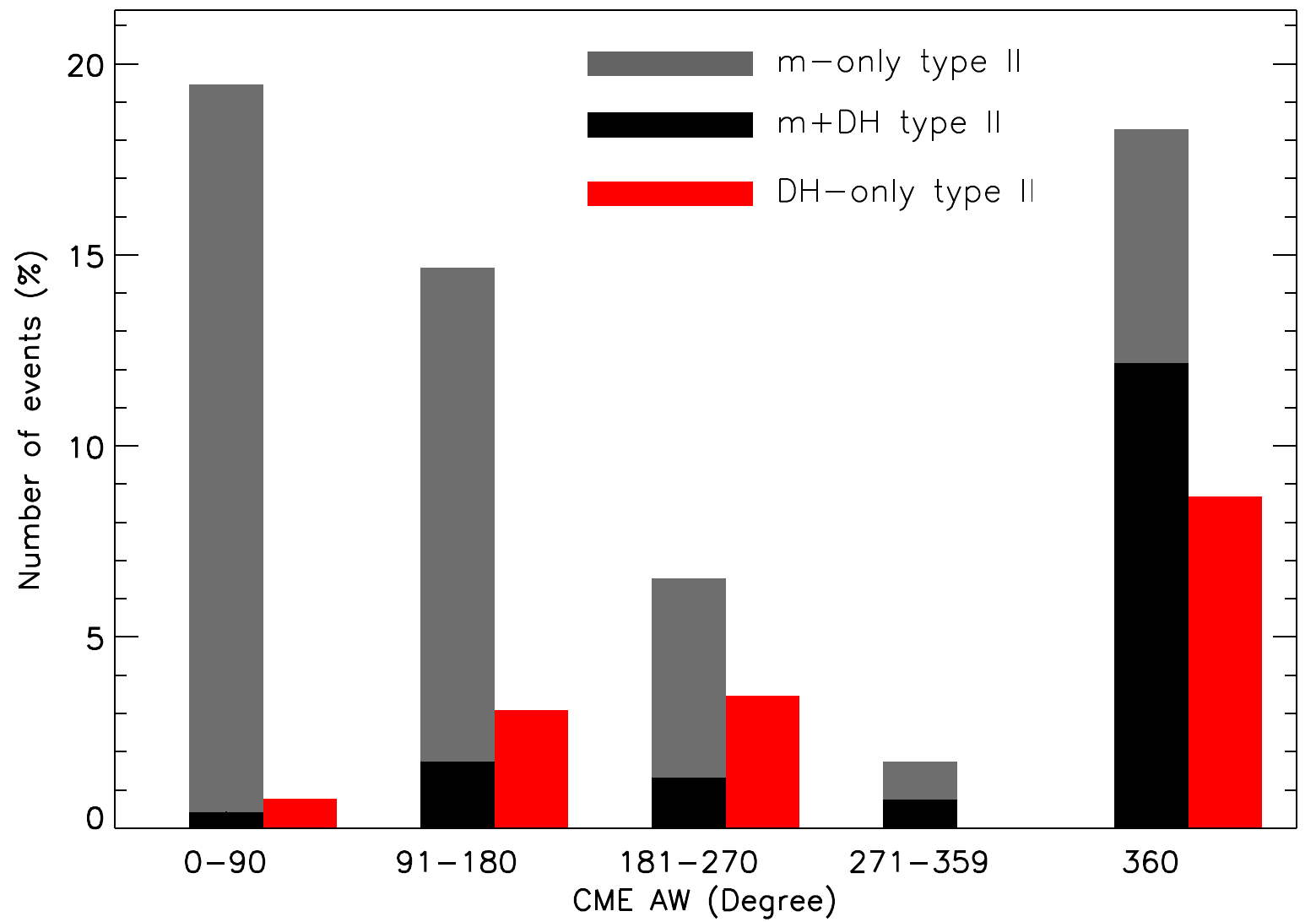}
    \caption{Histogram of the AW of the CMEs associated with m-only/m+DH/DH-only type II radio bursts. Color-code as in Figure~\ref{fig:year}.}
    \label{fig:CME-AW}
\end{figure}

From the distribution of AW (Figure~\ref{fig:CME-AW}) one can conclude that the majority of m-only IIs are associated with narrow CMEs: 19\% (99/518) have an AW in the range 0$-$90 degrees and 13\% (67/518) have AW of 91$-$180 degrees and their number continues to decline, however, there are 6\% (32/518) CMEs of 360 degrees AW (halo CMEs). From all type II-associated CMEs the fraction of the halos is 36\% (143/399), consistent with 6\% (32/518) m-only, 12\% (63/518) m+DH, and 9\% (45/518) DH-only IIs. When focusing on the fraction of halo CMEs within the sub-samples, however, we find 9\% (32/342) of the m-only, but 72\% (63/87) of the m+DH and 52\% (45/87) of the DH-only IIs to be associated with halo CMEs. In the study covering SCs 23 and 24 using the same data for DH type II radio bursts, \citet{Gopalswamy2019} found that more than half of the CMEs associated with DH type II radio bursts are halo CMEs. Similar results were reported by \citet{Reiner1998} and \citet{Gopalswamy2001}.

\subsubsection{Type II bursts and filament eruptions}

As discussed above, the type II radio bursts can be generated by the formation of shock at the CME front. The filament eruptions and CMEs are strongly associated with each other, for example, earlier studies reported this association to be more than 70\% \citep{Gopalswamy2003, Chen2011, Schmieder2013}. Therefore, the relationship between the type II radio bursts and filament eruptions is logical in case of filament eruptive CMEs. However, as per our knowledge, this relationship have not been reported in the literature till date.

Keeping this in view, we have investigated for the first time the relationship between type II radio bursts and filament eruptions. For this purpose, the data of filaments is taken from the provided links provided in Section \ref{data}. Out of the total number of type II bursts, 44\% (230/518) are associated with filament eruptions and 38\% (199/518) are not associated with filament eruptions. The remaining 17\% (89/518) type II radio bursts are the events for which the filament eruption data is not available. The distribution of m-only/m+DH/DH-only IIs associated filament eruptions are 25\% (132/518)/11\% (57/518)/8\% (41/518), respectively.
Subtracting the filament data gaps (89 cases) from the total number of events (518), we found $\sim$54\% (230/429) type II-associated filament eruptions. The weaker association of type II radio bursts and filament eruptions (54\%) in comparison to that with the CMEs (74\%) could be due to visibility issues of the filament detection using the current observational facilities.

\subsubsection{Type II Bursts and SN types} 

\begin{table}[t]
\caption{Table for the percentage of SN types associated with m-only/m+DH/DH-only II radio bursts given in \%, calculated as the ratio to entire list of 367 SF-associated type IIs. In parenthesis are given the exact number of events in each category. Here, `not reported' refers to the type II events which are associated with SFs but the SN type is not provided.}
\begin{tabular}{cccc}
\hline
\setlength{\tabcolsep}{0cm}
SN type &  \multicolumn{3}{c}{Association with type II} \\
(Magnetic Config) & m-only & m+DH & DH-only\\
\hline
$\alpha$ & 6.5\% (24) & 1.6\% (6) & 0.8\% (3) \\ 
$\alpha\gamma$ & 0.5\% (2) & 0\% (0) & 0.8\% (3) \\ 
$\alpha\gamma\delta$ & 0.3\% (1) & 0\% (0) & 0\% (0)\\ 
$\beta$ & 22.4\% (82) & 5.2\% (19) & 2.7\% (10) \\ 
$\beta\delta$ & 1.6\% (6) & 1.6\% (6) & 0.3\% (1) \\ 
$\beta\gamma$ & 10.4\% (38) & 4.9\% (18) & 2.7\% (10) \\ 
$\beta\gamma\delta$ & 10.6\% (39) & 6.0\% (22) & 2.7\% (10)\\ 
not reported & 10.4\% (38) & 2.2\% (8) & 3.3\% (12)\\ 
uncertain & 1.9\% (7) & 0.3\% (1) & 0\% (0)\\ 
\hline
    \end{tabular}
    \label{table1}       
    \end{table}

For this sub-section, we consider the list of 367 type II associated SFs and investigate their underlying SN configuration. Among these, 16\% (59/367) do not have a reported SN magnetic configuration and 2\% (8/367) have uncertain SN types, whereas the remaining 82\% (300/367) can be classified according to one of the Mount Wilson sub-categories. We examined the types of SNs and their occurrence rate with respect to m-only/m+DH/DH-only IIs. Due to the change in the sunspot configuration type during their lifetime, we investigate and report in Table~\ref{table1} the SN category present at the day of the type II burst occurrence. The percentages are calculated as ratio to 367 and in parenthesis are given the exact sample size.

The largest fraction of type II bursts are predominantly accompanied by activity from $\beta$ (30\%, 111/367), $\beta\gamma\delta$ (19\%, 71/367) and $\beta\gamma$ (18\%, 66/367) configuration, respectively. The exact distributions, also over the remaining SN configurations and the split into for m-only/m+DH/DH-only type II bursts, are shown in the Table~\ref{table1}.

\subsection{Type II bursts and space weather (SW) phenomena}

Similarly as above, we summarize the association rates between the SW phenomena and m-only/m+DH/DH-only II radio bursts, see Table~\ref{tab-SW}. There, only few, representative ranges (in particle flux, ICME/IP shock speed and GS Dst index) have been selected, and the details of the association rates can be inspected from the subsections below.

\begin{table}[t]
\caption{Table for the percentage of the solar phenomena associated with m-only/m+DH/DH-only II radio bursts given in \%, calculated as the ratio to entire list of 518 type IIs. SEP flux is in (cm$^{2}$ s sr MeV)$^{-1}$, SEE flux -- in (cm$^{2}$ s sr keV)$^{-1}$. In parenthesis are given the exact number of events in each category.}
\begin{tabular}{cccc}
\hline
\setlength{\tabcolsep}{0cm}
SW &  \multicolumn{3}{c}{Association with type II} \\
phenomena & m-only & m+DH & DH-only\\
\hline
SEP flux $0-0.2$ &  2\% (12)  &  5\% (25)  & 4\% (21) \\
SEP flux $>0.2$ &  0.4\% (2)  &  5\% (25)  & 1\% (7) \\ 
SEE flux $(0-1)\times 10^3$ &  6\% (30)  &  3\% (17)  & 4\% (21) \\
SEE flux $>1 \times 10^3$ &  1\% (8)  &  7\% (38)  & 2\% (11) \\
ICME$^W$ $\leq500$ \kms  &  4\% (23)  &  2\% (9)  & 0.2\% (1) \\
ICME$^W$ $>500$ \kms  &  0.4\% (2)  &  1\% (6)  & 2\% (12) \\
ICME$^A$ $\leq500$ \kms  &  2\% (10)  &  1\% (8)  & 2\% (13) \\
ICME$^A$ $>500$ \kms  &  6\% (30)  &  5\% (26)  & 0.8\% (4) \\
IP shock $\leq500$ \kms  &  6\% (34)  &  3\% (15)  & 2\% (12) \\ 
IP shock $>500$ \kms  &  2\% (11)  &  2\% (11)  & 0.9\% (5) \\
$|{\rm Dst}| \leq70$ nT &  3\% (15)  &  0.6\% (3)  & 0.6\% (3) \\
$|{\rm Dst}| >70$ nT &  0.8\% (4)  &  4\% (20)  & 2\% (11) \\
\hline
\end{tabular}
\label{tab-SW}       
\end{table}

\subsubsection{Type II bursts and in situ particles}

We explore the association between type IIs and in situ particles, which include solar energetic protons (SEPs) and electrons (SEEs) within an energy range 19$-$28 MeV and 103$-$175 keV, respectively. The overall occurrence rate of type II-associated SEPs is found to be 18\% (92/518) and the remaining 82\% type IIs are not associated with protons. Among the associations, 3\% (14/518) are m-only, 10\% (50/518) are m+IP and 5\% (28/518) are DH-only IIs related SEPs. 

\begin{figure}[t]
\includegraphics[width=0.9\columnwidth]{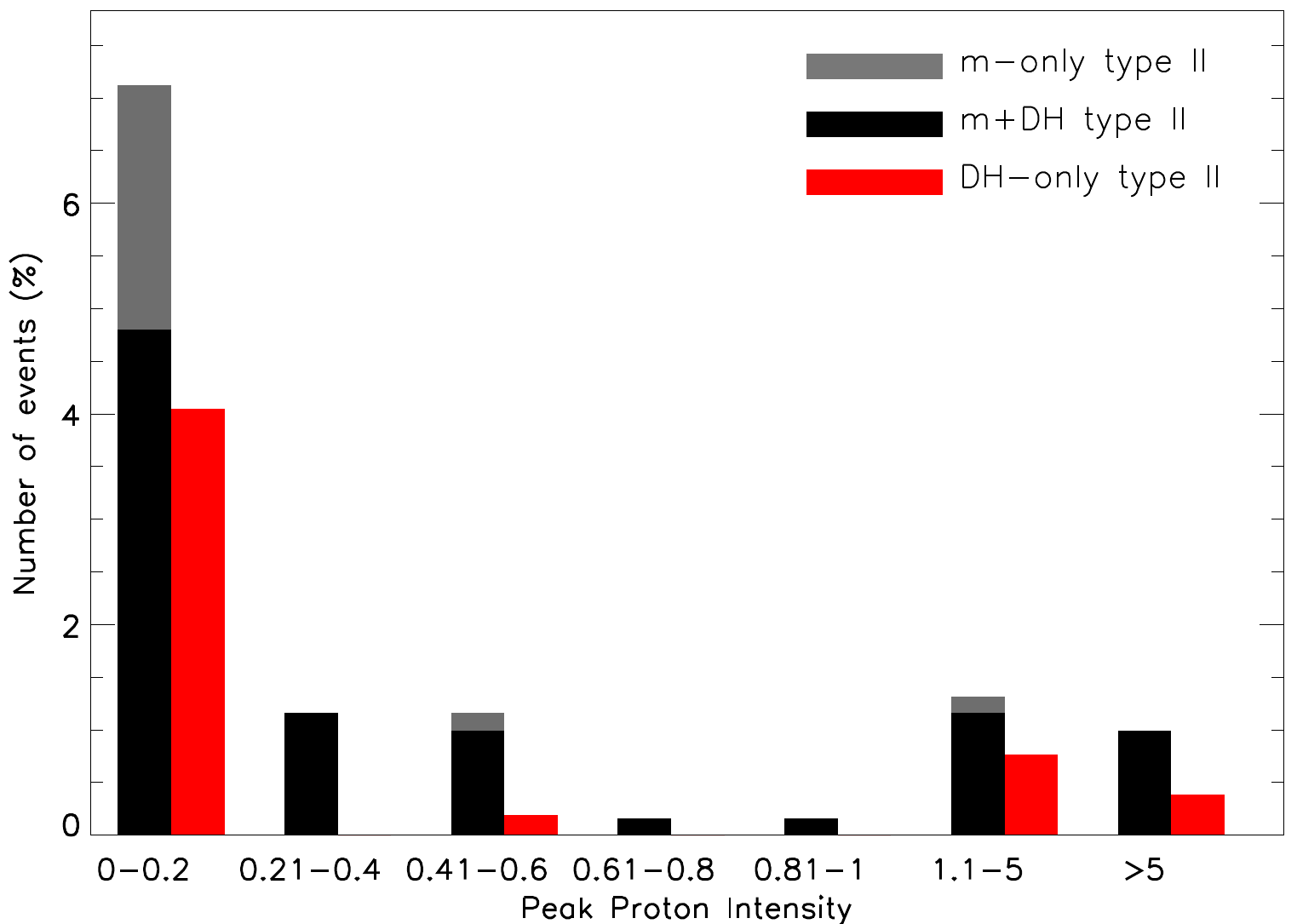}
\caption{Histogram of Wind/EPACT peak proton intensity of SEPs associated with m-only/m+DH/DH-only type II radio bursts. Color-code as in Figure~\ref{fig:year}.}
    \label{fig:SEPs}
\end{figure}

The distribution of the type II-associated SEPs as a function of their proton flux intensity is shown in Figure~\ref{fig:SEPs}. The majority type II associated SEP events have proton peak intensity $\leq$0.2 (cm$^{2}$ s sr MeV)$^{-1}$, for all categories of type IIs. Note that, this is also true for the entire SEP distribution in SC24, consisting of 142 events (not shown), as the weaker events constitute the majority of the population. The largest SEP values, however, are exclusively m+DH (and to a degree also DH-only) IIs associated protons. In case of DH-type II radio bursts, \citet{Winter2015} found that all the proton events $>$10 MeV (with flux $>$15 pfu (pfu = particle/(cm$^{2}$ s sr)) are associated with type II radio bursts in the period of 2010$-$2013.

Since the m-only IIs consists of 14 events, we calculate the mean, median, and peak SEP intensity for the entire sample of 92 events (of type IIs associated with SEPs) and obtain 4.86, 0.05, and 153 (cm$^{2}$ s sr MeV)$^{-1}$, respectively. For completeness, we calculated the same for the entire SEP sample (142 cases) and obtain 2.95, 0.03, and 153 (cm$^{2}$ s sr MeV)$^{-1}$, respectively.
Thus, the mean and median values of the type II-associated SEPs are nearly twice as large, compared to the entire type II related SEP population over SC24. 

\begin{figure}[t]
\includegraphics[width=0.9\columnwidth]{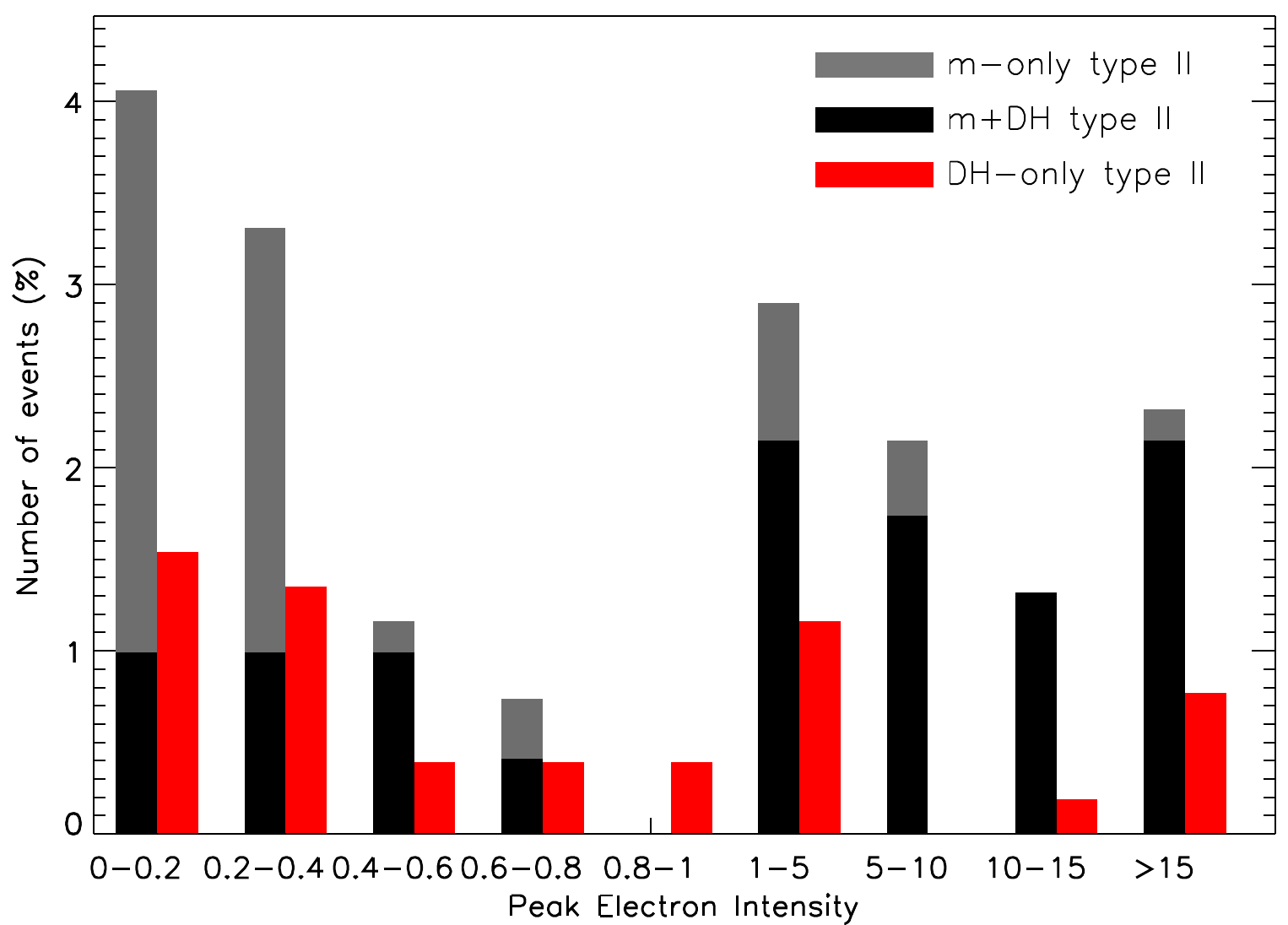}
\caption{Histogram of ACE/EPAM peak electron intensity of SEEs associated with m-only/m+DH/DH-only type II radio bursts. Color-code as in Figure~\ref{fig:year}.}
    \label{fig:SEEs}
\end{figure}

The distribution of peak electron intensity is shown in Figure~\ref{fig:SEEs} with the same colour notations as for Figure~\ref{fig:SEPs}. The occurrence rate for m-only, m+DH, and DH-only IIs associated SEEs is 7\% (38/518), 11\% (55/518), and 6\% (32/518), respectively, whereas the remaining type IIs are not associated with SEEs. The maximum number of SEE events is found again in the smallest intensity range $\leq$0.2$\times$10$^3$ (cm$^{2}$ s sr keV)$^{-1}$ but for the m-only IIs, whereas m+DH and DH-only IIs are more often associated with large SEEs (over a wider range of intensities, from 1 to over 15$\times$10$^3$). The mean, median, and peak electron intensity are:
\begin{itemize}
    \item for the entire list of 125 electron events related to type IIs: 13.7 $\times$ 10$^3$, 0.66 $\times$ 10$^3$, and 371 $\times$ 10$^3$ (cm$^{2}$ s sr MeV)$^{-1}$
    \item for 57 m+DH type IIs: 20 $\times$ 10$^3$, 4.2 $\times$ 10$^3$ and 371 $\times$ 10$^3$ (cm$^{2}$ s sr MeV)$^{-1}$,
\end{itemize}
whereas the m-only and DH-only have similar values at lower SEE intensity.

\subsubsection{Type II bursts and ICMEs}

For completeness, we present the association of type II radio bursts with ICMEs, as reported by two independent spacecraft, Wind (\url{https://wind.nasa.gov/ICME_catalog/ICME_catalog_viewer.php}) and ACE (\url{http://www.srl.caltech.edu/ACE/ASC/DATA/level3/icmetable2.htm}, \url{https://izw1.caltech.edu/ACE/ASC/DATA/level3/icmetable2.htm}), respectively.

\begin{figure}[t]
	\includegraphics[width=0.9\columnwidth]{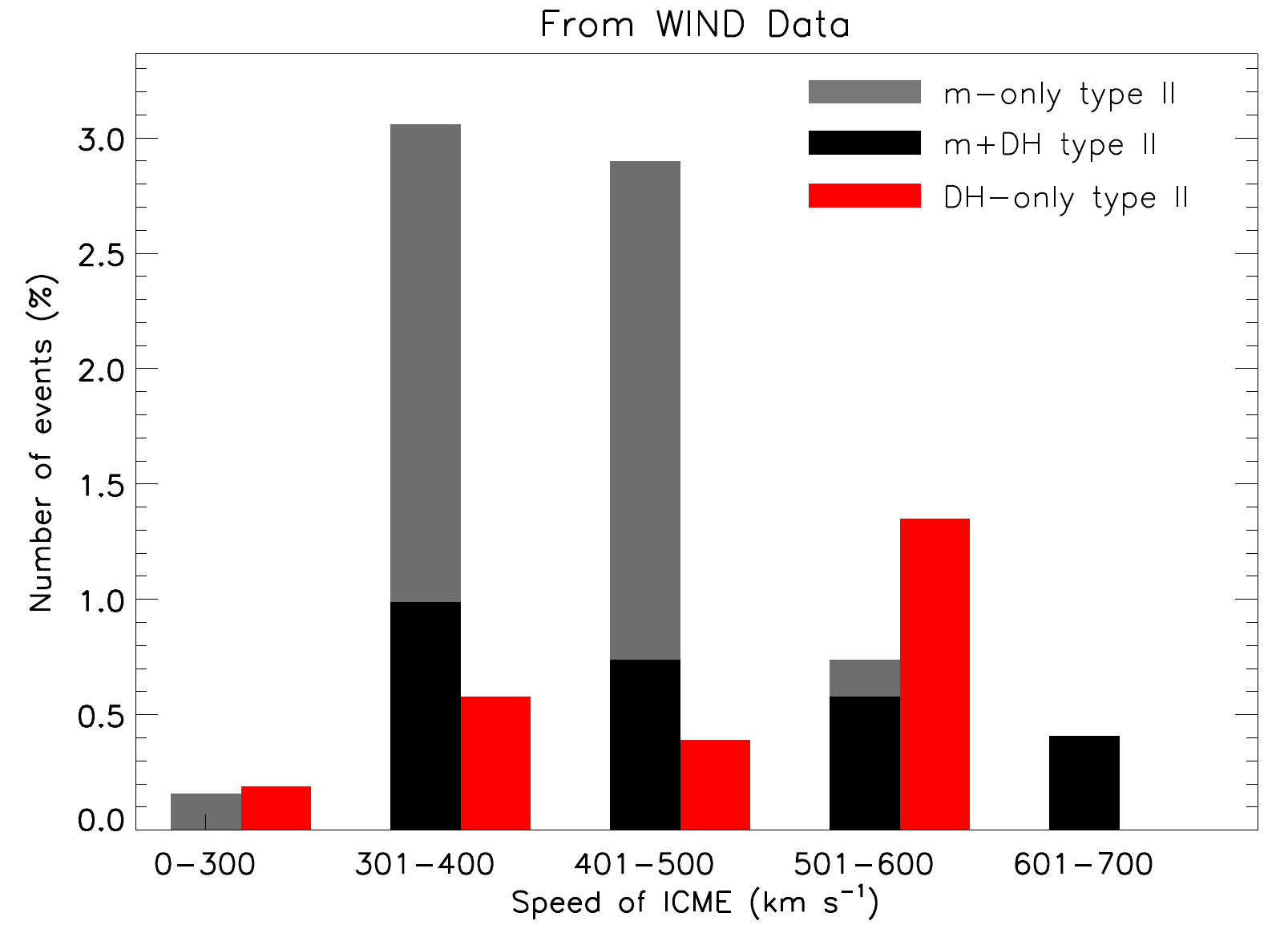}
    \caption{Histogram of the speed of ICMEs (Wind data) associated with m-only/m+DH/DH-only type II radio bursts. Color-code as in Figure~\ref{fig:year}.}
    \label{fig:ICME_Wind}
\end{figure}

We find a total of 10\% (53/518) type II-associated ICMEs as detected from the Wind spacecraft. Among them, m-only II-associated ICMEs are 5\% (25/518), m+DH are 3\% (15/518) and DH-only are 2\% (13/518), plotted in Figure~\ref{fig:ICME_Wind} as a function of the ICME speed. Most m-only type II events have speeds in the range 301$-$500 \kms. In contrast, in case of m+DH II-associated ICMEs are almost equally distributed in velocity in the range 301$-$700 \kms, wheres the DH-only II-associated ICMEs have larger speeds with the majority in the range 501$-$600 \kms. The mean and median values of the entire sample of type II-associated ICME speeds are 432 and 410 \kms, whereas the sample of all ICMEs detected by the Wind spacecraft have slightly slower mean/median speeds of 405/392 \kms, respectively.

\begin{figure}[t]
	\includegraphics[width=0.9\columnwidth]{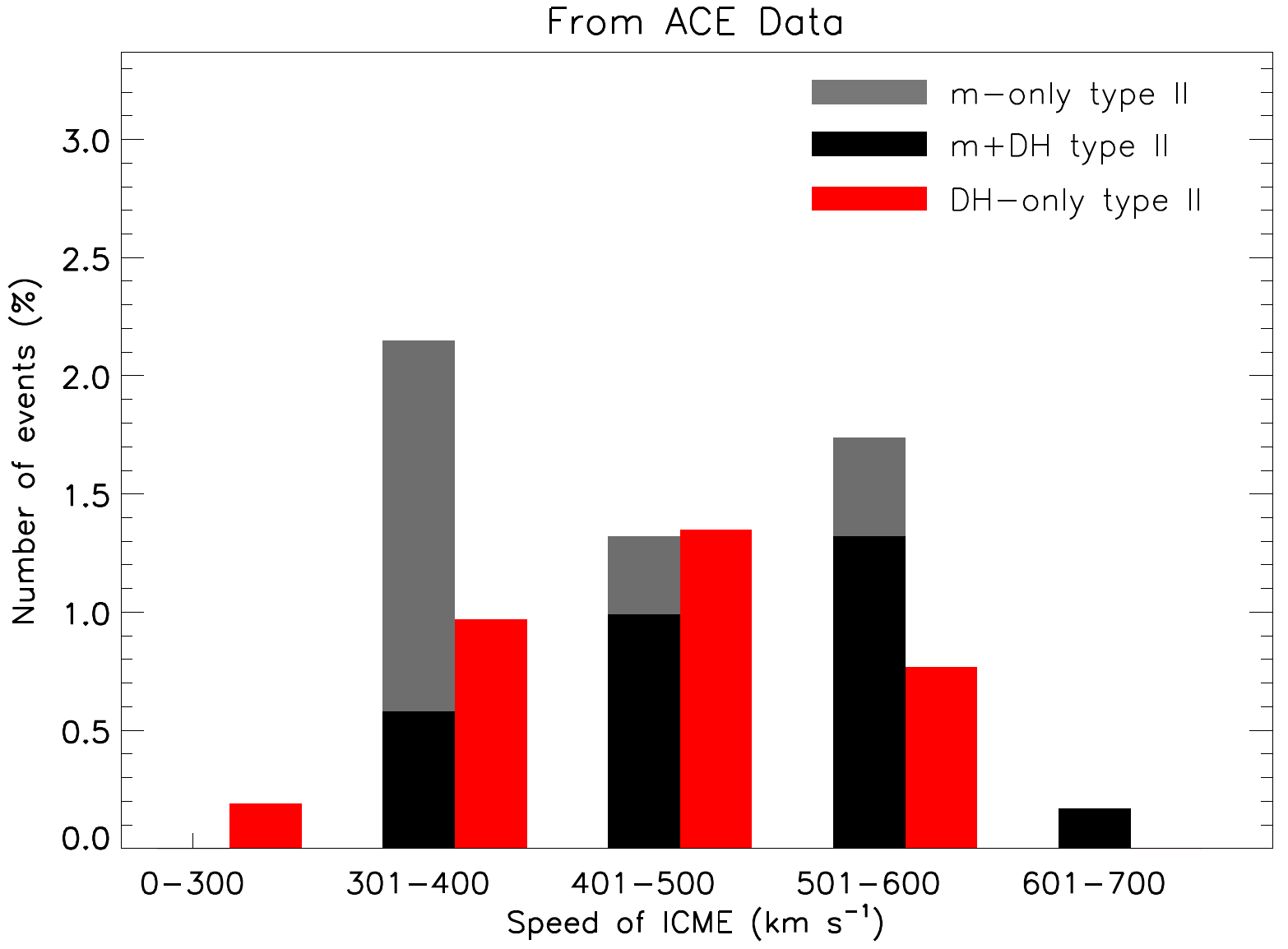}
    \caption{Histogram of the speed of ICMEs (ACE data) associated with m-only/m+DH/DH-only type II radio bursts. Color-code as in Figure~\ref{fig:year}.}
    \label{fig:ICME_ACE}
\end{figure}

Similarly, the type II-associated ICMEs observed by the ACE satellite are 18\% (91/518), divided into 8\% (40/518) m-only, 7\% (34/518) m+DH IIs and 3\% (17/518) DH-only IIs. Figure~\ref{fig:ICME_ACE} displays the speed distribution of type II-associated ICMEs as detected by the ACE spacecraft. Different trends are seen compared to the Wind results. Namely, ICMEs related to the m-only category are not abundant beyond 400 \kms. However, those associated with m+DH IIs peak at the 501$-$600 \kms, wheres DH-only IIs peak at lower speeds, 401$-$500 \kms. The discrepancy between Wind and ACE results can be explained by the different methodologies used to calculate the ICME speed. For m-only II associated ICMEs from ACE, the values for mean and median speeds are 460 and 455 \kms, respectively. Similarly, all ICMEs detected by ACE spacecraft have speeds of 420 and 410 \kms, respectively, which is consistent with the values obtained using Wind data.

Now we compare the fraction of ICME with the II radio bursts covering the DH range (m+DH and DH-only). Based on the ACE data with 91 ICMEs in SC24, we find 56\% (51/91) association between ICMEs and DH type II radio bursts. In case of Wind database this percentage is 53\% (28/53). 
Using the Richardson and Cane ICME catalog which is based on ACE satellite data, citet{Patel2022} did the analysis of DH type II and ICMEs for the SCs 23 and 24. They concluded the 47\% of association between them which is close to our association rate.

\subsubsection{Type II bursts and IP shocks}

We present the IP shocks as observed by the Wind satellite only, as ACE data is not available after 2013. In total we found about 17\% (89/518) type II associated IP shocks as reported by the Wind spacecraft for the SC24. These IP shocks are associated with 8\% (44/518) m-only, 5\% (26/518) m+DH and 4\% (19/518) DH-only IIs. 

\begin{figure}[t!]
	\includegraphics[width=0.9\columnwidth]{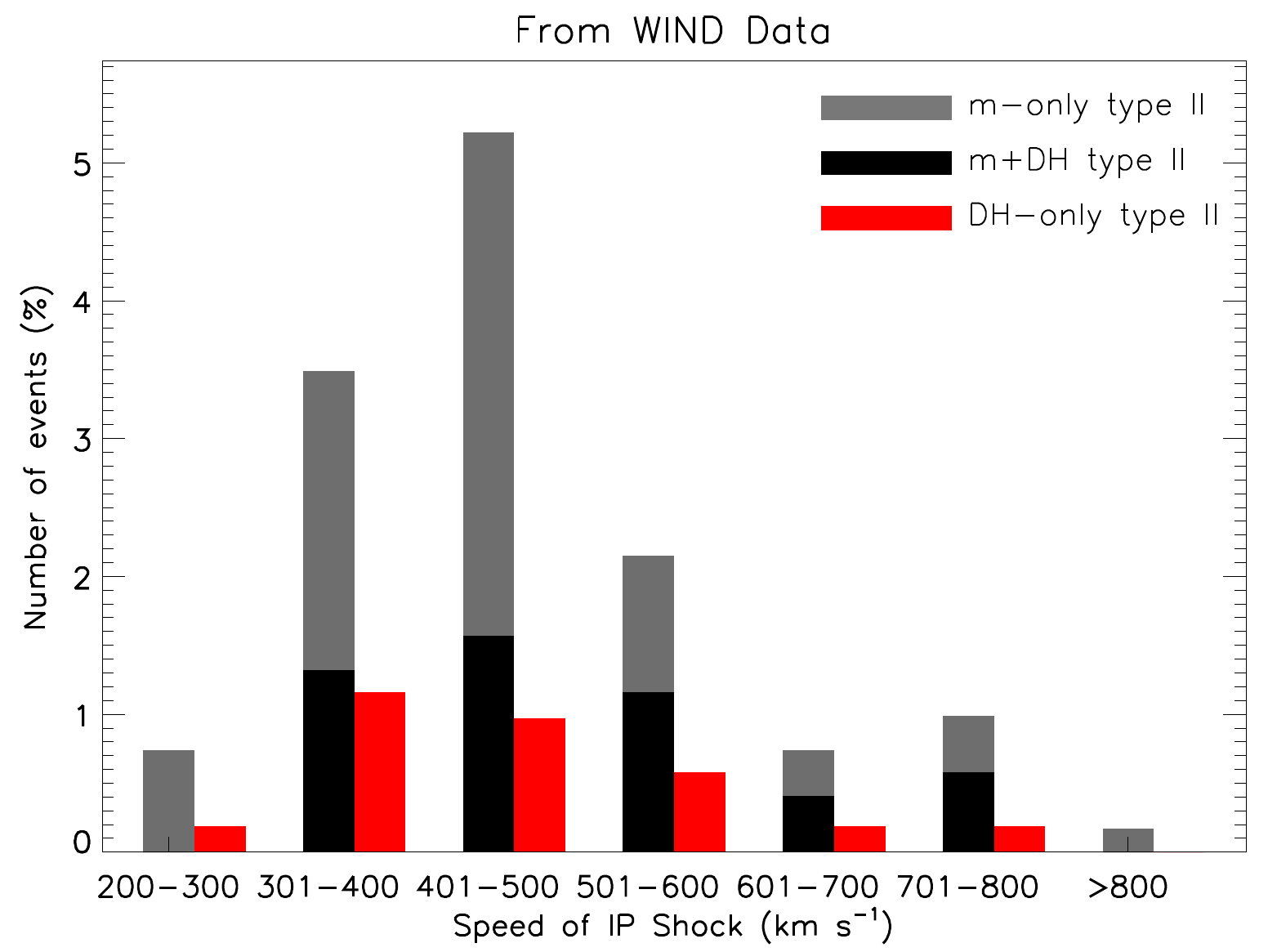}
    \caption{Histogram of the speed of IP shocks associated with m-only/m+DH/DH-only type II radio bursts. Color-code as in Figure~\ref{fig:year}.}
    \label{fig:IPshock}
\end{figure}

The distribution of speeds of IP shocks with respect to m-only, m+DH, and DH-only IIs is presented in Figure~\ref{fig:IPshock}. One could see that in terms of speed of the IP shocks, m-only and m+DH IIs are similarly distributed in the few bins, 301$-$600 \kms with a peak at 401$-$500 \kms. In contrast, the DH-only IIs peak at lower IP shock speeds. The mean/median values of the entire IP shock sample and the m-only II related IP shocks are found to be 421/412 and 465/444 \kms, respectively.

\subsubsection{Type II bursts and geomagnetic storms}

\begin{figure}
	\includegraphics[width=0.9\columnwidth]{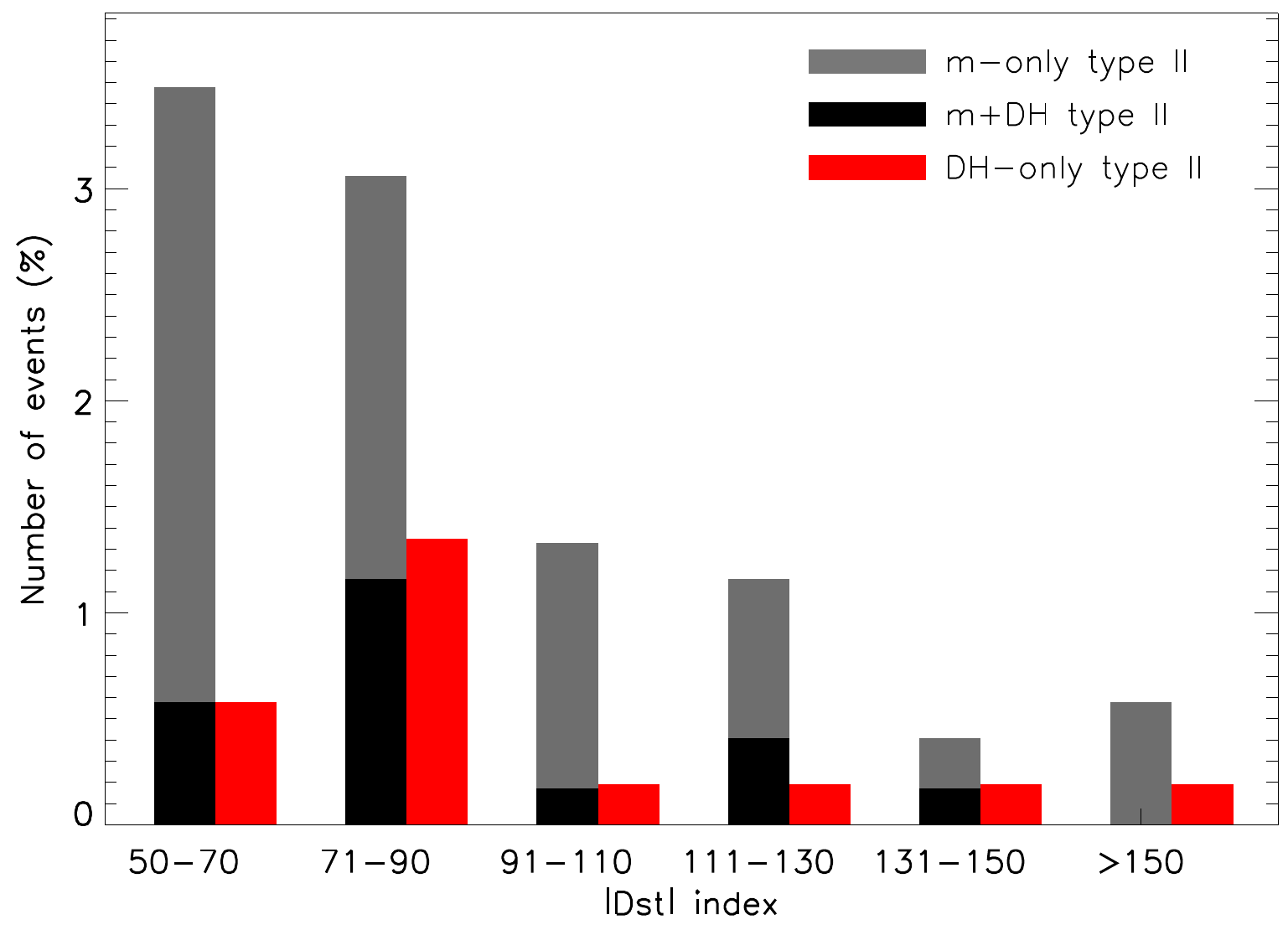}
    \caption{Histogram of the Dst index (in nT) of GSs associated with m-only/m+DH/DH-only type II radio bursts. Color-code as in Figure~\ref{fig:year}.}
    \label{fig:GS}
\end{figure}

We again start from type II radio bursts and this time we look for their association with GSs. We collect the data of GSs that occurred during SC24 with Dst index $\leq -50$~nT and the majority of them are weak storms. The data of Dst index is taken from Kyoto database. After the examination, 11\% (57/518) are found to be associated with type II radio bursts, among which 4\% (19/518) are m-only, 4\% (23/518) are m+DH and 3\% (15/518) DH-only IIs.

The distribution is in absolute values of the Dst index and is shown in Figure~\ref{fig:GS}. The majority of the m-only type II-associated GSs are distributed in the Dst range 50$-$70 nT (and also 71$-$90 nT), whereas m+DH and DH-only II-associated GSs show a peak at 71$-$90 nT range. The mean/median values of the m IIs (m-only and m+DH IIs) associated with GSs have $|{\rm Dst}|$ index 90/80.5 nT, respectively.


Furthermore, we examine in more details the CMEs associated with type II related GSs to look into the behaviour of GS producing CMEs. We find that out of all 57 GSs associated with our list of type II radio bursts
events, 7 are not associated with CMEs, and the remaining 47 have CMEs. The AW of these CMEs are mostly in range 60$-$119 (13 events i.e., 28\%) and 360 (16 events i.e., 34\%). Also, most of these CMEs have speeds in range of $\sim$ 200 to 1000 \kms.

\subsection{Overview on the association rates between type II bursts with solar and space weather events}

The information about the type IIs, as described in the study above, is summarized in a table-form for clarity (Table~\ref{tab-res}). Among the SN types, we selected to represent only the most abundant one, i.e., $\beta$. With respect to the ICMEs and IP shocks, only results based on the Wind data are listed, for consistency. In addition, we calculated the association rates between type IIs and GSs, the latter with Dst index of $\leq-100$ and $\leq-50$~nT (denoted as GS$^{-100nT}$ and GS$^{-50nT}$, respectively). However, GSs$^{-100\,nT}$ is not explicitly listed in the table due to the low number of events (13 (2.5\%) for the entire sample or 4/5/4 for m-only/m+DH/DH-only IIs).

\begin{table}[t!]
\centering
\caption{Table of the association between type IIs and SW events in SC24. The rates for the type IIs are given in \%, as a ratio to the number of all type IIs i.e., 518. The sample size in each case is given in parenthesis. (*Filament eruption association percentage is calculated by subtracting the 89 events for which filament eruption data is not available.)}
\medskip
\begin{tabular}{llllll}
\hline
\setlength{\tabcolsep}{20pt}
{\bf \# type IIs} & {\bf SFs} & {\bf CMEs} & {\bf $\beta$-type SNs} \\
\hline
518 type II   & 71\% (367) & 77\% (399) & 28\% (111)  \\
342 m-only II & 46\% (238) & 45\% (231) & 20\% (82) \\
87 m+DH II    & 15\% (80)  & 16\% (85)  & 5\% (19) \\
89 DH-only II & 9\% (49)   & 16\% (83)  & 3\% (10) \\
\hline
{\bf \# type IIs} &  {\bf SEPs} & {\bf SEEs} & {\bf filaments*} \\
518 type II       & 18\% (92)   & 24\% (125) & 44\% (230) \\
342 m-only II     & 3\% (14)    & 7\% (36)   & 25\% (132)  \\
87 m+DH II        & 10\% (49)    & 11\% (57)  & 11\% (57)   \\
89 DH-only II     & 5\% (28)    & 6\% (32)   & 8\% (41)  \\
\hline
{\bf \# type IIs} & {\bf ICMEs} & {\bf IP shocks} & {\bf GSs$^{-50\,nT}$}\\
\hline
518 type II    & 10\% (53) & 17\% (89) & 11\% (57) \\
342 m-only II  & 5\% (25)  & 8\% (44)  & 4\% (19) \\
87 m+DH II     & 3\% (15)  & 5\% (26)  & 4\% (23) \\
89 DH-only II  & 2\% (13)  & 4\% (19)  & 3\% (15)\\
\hline
\end{tabular}
\label{tab-res}       
\end{table}

\section{Discussion}
\label{discussion}

{\subsection{Association between type IIs and solar/SW phenomena}

In this study, we provide for the first time a statistical association between type II bursts in three wavelength ranges (m-only, m+DH, and DH-only) with a wide range of solar and IP phenomena over SC24. We point out that previous studies did not consider our strict sub-categorization and their event samples contain a mixture of m and DH type IIs. Furthermore, there is a tendency to focus more on the DH-type II due to their space weather relevance. Nevertheless, below we provide a comparative examination between our findings and earlier results.

Based on the results from this study (starting from type IIs and their association rates) one may argue that the overall radio signatures of shock waves are closely associated with solar eruptive phenomena, which are also their driver: 71\% (367/518) are associated with SFs, 77\% (399/518) with CMEs and 44\% (230/518) with filament eruptions (see also Table~\ref{tab-solar} for the sub-categories). \cite{Kumari2023} reported a 79\% association with CMEs in SC23, whereas in SC24 they claim 95\%. The difference stems from the adopted association criteria, level of uncertainty and/or the observer's subjectivity, as shown in \cite{2019BlgAJ..31...51M} for the case of SEPs and their solar origin. Furthermore, we investigated the SNs in terms of their magnetic morphology, and the majority of the type II bursts originated from $\beta$, $\beta$-$\gamma$ and $\beta$-$\gamma$-$\delta$ active region.

If we consider a physical effects of the SF eruption to the formation of shocks in the high corona and IP space (m+DH and DH-only II groups), these SFs tend to be stronger (M and X class) and of long duration. The magnetic configurations (Table~\ref{table1}) of the parent ARs and the latitudinal distributions of the SFs are in agreement with \citet{Gopalswamy2019} for DH type II radio bursts during SCs 23 and 24. We also confirm the well-known tendency of complex SN configurations to be related to eruptive events, e.g. \cite{2017MNRAS.465...68E}. 

Interestingly, the m+DH and DH-only II-accompanied CMEs also tend to be faster (over 500~\kms) and predominantly halo (namely, 72\% (63/87) of the m+DH and 52\% (45/87) of the DH-only IIs), compared to the slower and narrower CMEs accompanied the coronal shocks. These results are consistent with earlier reports: \cite{Gopalswamy2015CME} found that the CMEs with DH IIs have speed greater than 947 and 528 \kms~ during the two peaks of SC24; in a study covering SCs 23 and 24 using the same data for DH type II radio bursts, \citet{Gopalswamy2019} found that more than half of the CMEs associated with DH type II radio bursts are halo CMEs with similar results reported by \citet{Reiner1998} and \citet{Gopalswamy2001}. In summary, our findings agree with the notion that stronger drivers (e.g. faster CMEs) can lead to an IP phenomena as shown by \cite{Gopal2002,Prakash2014, Gopalswamy2015CME}. 

Using the onset type II frequency of the events from the RSTN data, reported originally by \cite{Lawrance2024}, we calculated here the mean (median) value of 30 (25) MHz for the m+DH II group compared to 44 (33) MHz for the m-only IIs, respectively. With the aid of a density model, these starting frequencies can be transformed into a coronal height indicative for the shock formation region. The results reveal that m+DH IIs are formed at predominantly higher layers and thus are able to propagate into the IP space and can have SW consequences.

With respect to SW phenomena, type II signatures show weaker associations, namely the type IIs can reach at most about 24\% association rates (i.e. with the electron events, Table~\ref{tab-res}). Despite the very low association rates, there are different trends for m-only vs. m+DH/DH-only IIs. For example, we obtained 56\% association rate between DH-only IIs and ICMEs, which is close to the 47\% reported by \citet{Patel2022} for DH type II and ICMEs in SCs 23 and 24. Furthermore, our findings show that the m+DH/DH-only IIs tend to be accompanied by in situ particle events of larger flux, faster ICMEs (based on Wind data) and stronger GSs (Table~\ref{tab-SW}), although due to the small event samples the results are mostly not (or only marginally) statistically significant. Collecting data over the ongoing SC25 is expected to improve the overall statistics and validate the obtained trends.

\citet{Prakash2014} studied the geo-effectiveness and type II associated CMEs during the period 1997$-$2005 (SC23). They concluded that 92\% of geoeffective CMEs are associated with DH type II radio bursts. In order to compare with the our findings we focus on our sub-sample of 85 m+DH II associated CMEs and found that only 15\% (13/85) are associated to GSs. Namely, the majority of GSs are associated with DH type II radio bursts whereas a small fraction of DH type II radio bursts can be associated to GSs.

\subsection{Association between solar/SW phenomena and type IIs}

The above results are based on the associations done when starting with a list of type IIs and exploring the respective solar and IP signatures. The opposite direction of association shows stronger occurrence rates between the solar or/and SW phenomena in SC24 and type IIs, although previous studies did not discriminate into m-only/m+DH/DH-only IIs but instead reported on the more generic m vs. DH type IIs. The results are listed below:
\begin{itemize}
    \item SFs: M-class to DH IIs, 7\% \citep{Miteva2022a}; X-class to m IIs, 67\%; X-class to DH IIs, 45\% (this study)
    \item CMEs: to m IIs, 3\% \citep{Kumari2023}
    \item SEPs: to m IIs, 59\%; to DH IIs, 73\% \citep{Miteva2017} 
    \item SEEs: to m IIs, 25\%; to DH IIs, 29\% \citep{Miteva2022}
    \item ICMEs: to DH IIs, 47\% \citep{Patel2022}
    \item IP shocks: to DH IIs, 95\% (upper limit, this study)
    \item GSs: $-50$ nT GSs to DH IIs, 19\%; $-100$ nT GSs to DH IIs, 38\% (this study)
\end{itemize}

{\bf \subsection{Type IIs and SW forecasting}}

Thus, despite the limited correspondence between an observed type II bursts and a SW event, the IP disturbances often follow a (DH) type II. Due to the faster arrival at Earth of the solar particles (of the order of minutes for the SEEs to hours for SEPs), the use of type II signatures for their timely forecasting has limited potential \citep{Nunez2020}. The latter empirical model relies on near-real time reports of radio emission signatures and the time needed for the development of the type IIs in the corona and the subsequent propagation and detection in the IP space adds a substantial delay. The arrival of the ICMEs, IP shocks and their subsequent GS effects, however, may take days after the parent solar eruptions. Thus, following the development of the type II signatures from m to DH ranges could be important in terms of their SW forecasting (e.g. using probabilistic techniques) and deserves further investigation \citep{Fry2003, 2007SpWea...5.8001C}. The reported occurrence rates and association of type IIs with solar and IP phenomena together with the previously estimated physical parameters by \cite{Lawrance2024} can be utilized for the improvement of physics-based models on synthetic type II radio signatures as discussed in \cite{2021A&A...654A..64J} due to the large type II event sample reported by us.

The use of a type II radio signatures in empirical or physics-based models of different SW phenomena was summarized early on in the review by \cite{2008A&ARv..16....1P}. More recently, metric type IIs were included in neural network techniques for the ICME arrival forecasting \citep{2019Ap&SS.364..161N} and the use of DH type IIs in empirical forecasting models was demonstrated by \cite{2023SoPh..298..145M} for the case of shock arrival at Earth. Our event sample not only covers the entire SC24 but is already sub-divided into specific, coronal and IP, categories, and thus, the aforementioned existing models can be readily re-examined by using our event samples as input parameters. Our type II catalog together with all associated solar and SW events will be made freely accessible and supported in the future via \url{https://catalogs.astro.bas.bg/}.

\section{Summary}
\label{summary}

In this study, we report on the association between type II radio bursts, as observed by a chain of four different radio stations (i.e., RSTN) with selected solar and SW events during 2009$-$2019 i.e. SC24. We started with a list of coronal type IIs and added the reported type IIs in the IP space. The association rates between the type II radio bursts (m-only/m+DH/DH-only IIs) and the solar activity and SW phenomena are summarized below:

\begin{itemize}
\item {The majority of the radio bursts in our list, 66\% (342/518), occur only in the solar corona (m-only IIs). About 17\% (87/518) are found to be accompanied with DH type IIs (m+DH) implying that these coronal bursts become IP phenomena and the same percentage (or 89/518) is observed only in the IP space (DH-only IIs).}
    
\item {The type II associated SFs amount to 71\% (367/518) of the entire radio burst sample. The majority of these SFs have C and M-classes (m-only and DH-only II category, whereas m+DH II-associated SFs are mostly of M and X-class). The m-only IIs are impulsive (i.e., have short rise times 6$-$15 mins), whereas SF-associated m+DH type IIs show no clear impulsive trend and DH-only IIs are related to long-duration SFs. No clear longitudinal dependence is noticed, whereas in helio-latitude the SFs tend to originate around the 10$-$30-degree AR belts, irrelevant on their m-only/m+DH/DH-only type II association.}
    
\item {Out of the total type II radio bursts, $\sim$77\% are associated with CMEs (399/518), distributes into 45\%/16\%/16\% for m-only/m+DH/DH-only IIs, respectively. The m-only type II-associated CMEs predominantly have speeds below 500 \kms and are narrow, with AW below 180 degrees. In contrast, the CMEs associated with m+DH and DH-only IIs are faster (mostly in the range 501$-$1500 \kms) and exclusively halo.}

\item {The magnetic configurations of the SNs are examined and it is found that 82\% of the type IIs can be classified in terms of sunspot type with the majority (30\%) being $\beta$-type.} 

\item {There is a weak association between type IIs and in situ particles, in total of 17\% with SEPs and 24\% with SEEs. In both cases, the largest fractions are obtained for the m+DH type IIs, 9\% and 11\%, whereas the remaining m-only and DH-only group are of the order of few percents.}
        
\item {About 44\% of all type IIs are found to be clearly associated with filament eruptions, excluding the data gaps, distributed as 25\%/11\%/8\% to m-only/m+DH/DH-only IIs, respectively.}

\item {The type IIs related to ICMEs from Wind and ACE spacecrafts are 11\% and 18\%, respectively. The distribution into the three subcategories is nearly the same for the Wind data, whereas ACE data gives larger association rates with m-only and m+DH type IIs.}

\item {The type IIs related to IP shocks based on data from the Wind satellite amount to 17\% with similar distribution as the ACE ICMEs.}
    
\item  {The association rate of GSs (Dst index $\leq - 50$~nT) with type IIs is found to be $\sim$11\%, with similar distribution as for the ICMEs.} 
\end{itemize}

\section*{Appendix: Wind/WAVES DH IIs and associated SW phenomena}

\begin{table}[t!]
\centering
\caption{Table of the association between DH IIs and SW phenomena, in three different time periods, given as a ratio to their total number depicted in the respective column. The sample size in each case is given in parenthesis.}
\medskip
\begin{tabular}{llll}
\hline
\setlength{\tabcolsep}{20pt}
{Events}       & SC23  & SC24  & SC23+24\\
                   & (341) & (181) & (522) \\
\hline
{SFs}          & 88\% (299) & 61\% (111) & 77\% (410) \\
{CMEs}         & 96\% (326) & 96\% (174) & 96\% (500) \\
{$\beta$-type SNs} & 23\% (79)  & 15\% (28)  & 20\% (107) \\
{SEPs}         & 36\% (122) & 41\% (75)  & 38\% (197) \\
{SEEs}         & 20\% (151) & 19\% (84)  & 20\% (235) \\
{ ICMEs}        & 19\% (62)  & 13\% (23)  & 16\% (85) \\
{GSs$^{-100\,nT}$} & 7\% (24) & 12\% (21) & 9\% (45) \\
{GSs$^{-50\,nT}$} & 28\% (95) & 20\% (37) & 25\% (132) \\
\hline
\end{tabular}
\label{tab-App-A}       
\end{table}

For completeness, we calculated the associations between DH IIs (reported by \url{https://cdaw.gsfc.nasa.gov/CME_list/radio/waves_type2.html}) and the considered above SW events, see Table~\ref{tab-App-A}. All associations are calculated when starting from a list of DH IIs. The lists of SW events are the same used in the above analyses. Only due to the lack of catalogs of filaments and IP shocks with already identified solar origin (SFs and/or CMEs), we cannot present their occurrence rates with the DH II samples and thus these events are dropped. Note that the reported DH IIs cover both m+DH and DH-only categories from our analyses, but also extend to the previous SC23.

The number of DH II bursts are 341 in SC23, 181 in SC24 or in total of 522 DH II events in SC23+24. All occurrence rates are shown in \% and normalized to the latter sample sizes. Due to instrumental data gaps, the reported associations above should be considered as lower limits only. 

Firstly, we preformed the association between the DH IIs and the SFs and/or CMEs. In SC23 for the SFs we obtain 88\% (299/341), in SC24 61\% (111/181) or overall 77\% (410/522). The association rates for CMEs is the same (96\%) in either time period, confirming the well-known strong correlation between DH IIs and CMEs.

Similarly, we identified the occurrence of $\beta$-type SNs magnetic configurations, SEPs, SEEs and ICMEs with respect to the DH II, based on the same SF or CME for either pair. The associations vary between 10$-$20\% (for ICMEs and $\beta$-type SNs) to about 40\% (for the SEPs). 
 
Much weaker occurrence is calculated for the DH IIs and GSs. For the comparison with GSs we use the recently released catalogs of larger GSs with
$|{\rm Dst}|>$ 100~nT (denotes as GSs$^{-100nT}$) and an extended version, including weaker GSs with $|{\rm Dst}|>$ 50~nT (denotes as GSs$^{-50nT}$). Both catalogs provide the solar origin of these GSs which are compared with the solar origin of the DH IIs. The associations with larger GSs is around 10\%, whereas the GSs including the weaker storms increases from 20\% in SC24 to nearly 30\% in SC23.
\\

\noindent {\it Acknowledgments:}
The authors thank the reviewers for the comments and suggestions and the open data policy of RSTN, GOES, SOHO, ACE and Wind instruments, as well as the WDC-SILSO (Royal Observatory of Belgium) and WDC for Geomagnetism (Kyoto) databases. This research was supported by the Bulgarian-Indian Project KP-06-India/14 funded by the National Research Fund of Bulgaria and India. PD acknowledges the support from CSIR, New Delhi. RC acknowledge the support from DST/SERB project number EEQ/2023/000214.

\bibliographystyle{model5-names}
\biboptions{authoryear}
\bibliography{reference}

\end{document}